\documentstyle[epsfig,12pt]{article}

\def\ga{\mathrel{\mathpalette\fun >}}
\def\fun#1#2{\lower3.6pt\vbox{\baselineskip0pt\lineskip.9pt
\ialign{$\mathsurround=0pt#1\hfil##\hfil$\crcr#2\crcr\sim\crcr}}}

\title{CKM matrix and CP violation in B-mesons}
\author{M. Vysotsky \\ ITEP, 117218, Moscow, Russia}
\date{}
\begin{document}

\maketitle

\begin{abstract}

Planned originally as a review of CP violation (CPV) in B-mesons
which covered recent B-factories results these lectures turned out
to be a bit wider. It is not natural to be limited by CPV in
decays and mixings of B-mesons and not to speak about the
analogous phenomena in K-mesons since it is very useful and
interesting to study what is common and what is different in these
systems and why. CKM matrix elements are extracted from K and B
mixings and decays and the deviation from unitarity of CKM matrix
may become the place in which New Physics will show up. So we
discuss this simple and elegant piece of Standard Model as well.

In order to follow these lectures you should be able to write
the Lagrangian and to draw the Feynman diagrams in a Standard Model
and to calculate the corresponding amplitudes.

\end{abstract}

\newpage

\begin{center}

{\bf Plan of the lectures}

\end{center}

\begin{description}
\item{1.} ``The road map''
\item{2.} CKM matrix -- where from?
\item{3.} CKM matrix: angles, phases, parametrization, unitarity
triangles
\item{4.} $V_{us}$, $V_{cb}$, $V_{ub}$ -- {\it first circle}
\item{5.} CPV: history; why  phases are relevant
\item{6.} $M^0$ -- $\bar M^0$ mixing, CPV in mixing
($|\frac{q}{p}|\neq 1$)
\item{7.} Space-time pattern of $K^0$ -- $\bar K^0$ ($B^0$ --
$\bar B^0$, $\nu_\mu$ -- $\nu_e)$ oscillations
\item{8.} $K^0$ -- $\bar K^0$ mixing, $\Delta m_{LS}$
\item{9.} CPV in $K^0$ -- $\bar K^0$ mixing, $K_L \to 2\pi$ decay,
$\varepsilon_K$ -- {\it hyperbola}
\item{10.} Direct CPV in K decays, $\varepsilon^\prime \neq 0$
($|\frac{\bar A}{A}|\neq 1$)
\item{11.} $B_d^0$ -- $\bar B_d^0$, $B_s^0$ -- $\bar B_s^0$ mixings -- {\it two
circles}
\item{12.} CPV in $B^0$ -- $\bar B^0$ mixing, $a_{SL}^{(B_q)}$ -- too small
effects
\item{13.} CPV in interference of mixing and decay
($Im(\frac{q}{p}\frac{\bar A}{A}) \neq 0$)
\item{14.}
$B_d^0 (\bar B_d^0) \to  J/\psi K$, $\sin 2\beta$ -- {\it straight
lines}
\item{15.} $B \to \pi\pi$, $\sin 2\alpha$, penguin versus tree,
$|\frac{\bar A}{A}| \neq 1$
\item{16.} Angle $\gamma$
\item{17.} CPV in $B \to \phi K_S$, $K^+ K^- K_S$, $\eta^\prime K_S$: penguin
domination
\item{18.} Conclusions: CKM fit and future prospects

\end{description}

\newpage

\section{``The road map''}

In Fig.1 you can see a set of bounds on the parameters
$\bar{\rho}$ and $\bar{\eta}$ ($\bar{\rho}$ and $\bar{\eta}$ are
defined by eq.(3.10)) of the Cabibbo-Kobayashi-Maskawa \cite{1,2}
quark mixing matrix (CKM). They comprise three circles, two
branches of a hyperbola, and two straight lines. Three circles
originate from the $V_{ub}$ (for definition of $V_{ub}$ see
eq.(3.4)) measurement (the green one with the center at
$\bar{\rho} = \bar{\eta} =0$,
 see eq.(4.7)), the measurement of $\Delta m_{B_d}$ -- the mass difference of two
 energy eigenstates of $B_d$ and $\bar B_d$ mesons (the red one, eq.(11.16) with
 the center at $\bar\rho =1$, $\bar\eta =0$)
 and from the lower bound on $\Delta m_{B_s}$ which is the same for $B_s$ and $\bar B_s$
mesons (the yellow one, eq.(11.20)). The hyperbola originates from
the measurement of CP violation -- CPV -- in the mixing of
$K$-mesons (see eq.(9.9)). Straight lines come from the
measurement of CP asymmetry in $B_d^0(\bar B_d^0) \to J/\Psi K$
decays (see eq.(14.11)). The fact that all  three circles,
hyperbola and straight lines intersect at one and the same place
means the triumph of Standard Model. Our aim is to explain in
these lectures where all these bounds come from.

\begin{figure}[!htb]
\centering
\epsfig{file=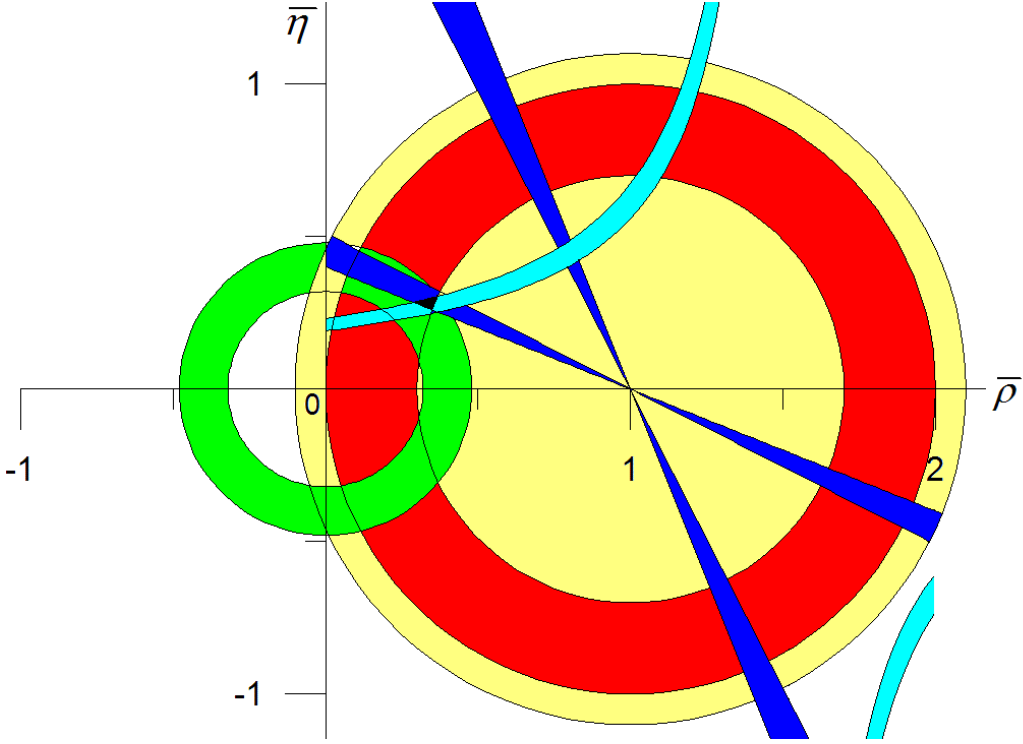,width=10cm,angle=90}
\caption{\em  The domains at $(\bar{\rho}, \bar{\eta})$ plane allowed at $1 \sigma$ from $V_{ub}$, $\Delta m_{B_d}$,
$\varepsilon_K$ and $\sin 2\beta$ measurements. 95{\rm \%}C.L. upper bound
from the search of $\Delta m_{B_s}$ is shown as well.}
\label{WW1Fermi}
\end{figure}

\section{CKM matrix -- where from?}

In constructing the Standard Model Lagrangian the basic ingredients
are 1. gauge group, 2. particle content and 3. renormalizability of the
theory. There is no such a building block in Standard Model as CKM
matrix in charged currents quark interactions. CKM matrix
originates from Higgs field interactions with quarks. The piece of
the Lagrangian from which the up quarks get their masses looks
like: $$ \Delta{\cal L}_{\rm up} = f_{ik}^{(u)} \bar
Q_L^{i^\prime} u_R^{k^\prime} H + {\rm c.c.} \; , \;\; i,k = 1,2,3
\;\; , \eqno(2.1) $$ where $$ Q_L^{1^\prime} = \left(
\begin{array}{cc} u^\prime
\\ d^\prime \\ \end{array} \right)_L \; , \;\; Q_L^{2^\prime} = \left( \begin{array}{cc}
c^\prime \\ s^\prime \\ \end{array} \right)_L \; , \;\;
Q_L^{3^\prime} = \left(
\begin{array}{cc} t^\prime \\ b^\prime \\ \end{array} \right)_L \;\; ; $$
$$u_R^{1^\prime} = u_R^\prime \; , \;\; u_R^{2^\prime} =
c_R^\prime \; , \;\; u_R^{3^\prime} = t_R^\prime \eqno(2.2) $$ and
$H$ is the Higgs doublet: $$ H = \left(\begin{array}{cc} H^0
\\ H^-
\end{array}\right)
\eqno(2.3)
$$

The piece of the Lagrangian which is responsible for the down
quark masses looks the same way: $$ \Delta{\cal L}_{\rm down} =
f_{ik}^{(d)} \bar Q_L^{i^\prime} d_R^{k^\prime} \tilde H + {\rm
c.c.} \; , \eqno(2.4) $$ where $$ d_R^{1^\prime} = d_R^\prime \; ,
\;\; d_R^{2^\prime} = s_R^\prime \; , \;\; d_R^{3^\prime} =
b_R^\prime \;\; {\rm and} \;\; \tilde H_a = \varepsilon_{ab} H_b^*
\; , \eqno(2.5) $$ $$ \varepsilon_{ab} = \left( \begin{array}{rl}
0 & 1
\\ -1 & 0
\end{array} \right) \; .
$$

After $SU(2) \times U(1)$ symmetry breaking by the Higgs field
expectation value  $<H^0> = v$ from formulas (2.1) and (2.4) two
mass matrices emerge: $$ M_{\rm up}^{ik} \bar u_L^{i^\prime}
u_R^{k^\prime} + M_{\rm down}^{ik} \bar d_L^{i^\prime}
d_R^{k^\prime} + c.c. \eqno(2.6) $$

The matrices $M_{\rm up}$ and $M_{\rm down}$ are arbitrary
3$\times$3 matrices; their matrix elements are complex numbers.
According to the well-known theorem an arbitrary matrix can be
written as a product of the hermitian and unitary matrices: $$ M =
UH \; , \;\; {\rm where} \;\; H = H^+ \;, \;\; {\rm and} \;\; UU^+
=1 \;\; , \eqno(2.7) $$ (do not mix the hermitian matrix $H$ with
the Higgs field doublet) which is analogous to the following
representation of an arbitrary complex number:  $$a = e^{i\phi}
|a| \;\; . \eqno(2.8) $$ From eq. (2.7) it is evident that matrix
$M$ can be diagonalized by 2  different unitary matrices acting
from left and right: $$ U_L M U_R^+ = M_{\rm diag} =
\left(\begin{array}{ccc} m_u & & 0 \\ & m_c & \\ 0 & & m_t
\end{array} \right) \;\; , \eqno(2.9) $$
where $m_i$ are real numbers (if matrix $M$ is hermitian ($M =
M^+$) then we will get $U_L = U_R$). Having these formulas in mind
let us rewrite the up-quarks mass term from eq. (2.6): $$ \bar
u_L^{i^\prime} M_{ik} u_R^{k^\prime} + c.c. \equiv \bar u_L^\prime
U_L^+ U_L M U_R^+ U_R u_R^\prime + c.c. = \bar u_L M_{\rm diag}
u_R^+ + c.c. = \bar u M_{\rm diag} u \;\; , \eqno(2.10) $$ where
we introduce the fields $u_L$ and $u_R$ according to the following
formulas: $$  u_L = U_L u_L^\prime \; , \;\; u_R = U_R u_R^\prime
\;\; . \eqno(2.11) $$

Applying the same procedure to matrix $M_{\rm down}$ we observe
that it becomes diagonal as well in the rotated basis: $$ d_L =
D_L d_L^\prime \; , \;\; d_R = D_R d_R^\prime \;\; . \eqno(2.12)
$$

Thus we start from the primed quark fields and get that they
should be rotated by 4 unitary matrices $U_L$, $U_R$, $D_L$ and
$D_R$ in order to obtain unprimed fields with diagonal masses.
Since kinetic energies and interactions with the vector fields
$A_\mu^3$, $B_\mu$ and gluons are diagonal in the quark fields,
these terms remain diagonal in a new unprimed basis. The only term
in the SM Lagrangian where matrices $U$ and $D$ show up is charged
current interaction with the emission of $W$-boson: $$ \Delta{\cal
L} = g W_\mu^+ \bar u_L^\prime \gamma_\mu d_L^\prime = gW_\mu^+
\bar u_L \gamma_\mu U_L^+ D_L d_L \;\; , \eqno(2.13) $$ and the
unitary matrix $V\equiv U_L^+ D_L$ is called
Cabibbo-Kobayashi-Maskawa quark mixing matrix.

\section{CKM matrix: angles, phases, parametrization, unitarity
triangles}

One can easily check that $n\times n$ unitary matrix has $n^2/2$
complex or $n^2$ real parameters. The orthogonal $n\times n$
matrix is specified by $n(n-1)/2$ angles (3 Euler angles in case
of $O(3)$). That is why the parameters of the unitary matrix are
divided between phases and angles according to the following
relation: $$
\begin{array}{cccc} n^2 = & {\frac{n(n-1)}{2}} & + &
\frac{n(n+1)}{2} \;\; . \\ &&& \\ & {\rm angles}& & {\rm phases}
\end{array} \eqno(3.1) $$

The next question arises: are all these phases physical
observables or, in other words, can they be measured
experimentally. And the answer is ``no'' since we can perform
phase rotations of quark fields ($u \to e^{i\zeta}u$, $d\to
e^{i\xi} d$ ...) removing in this way $2n-1$ phases of the CKM
matrix. The number of unphysical phases equals that of up and down
quark fields minus one since the simultaneous rotation of all
up-quarks on one and the same phase multiplies by (minus) this
phase all the matrix elements of matrix $V$. The rotation of all
down-quark fields on one and the same phase acts on $V$ in the
same way. That is why the number of the ``unremovable'' phases of
matrix $V$ is diminished by the number of possible rotations of up
and down quarks minus one.

Finally for the number of observable phases we get:
$$\frac{n(n+1)}{2} - (2n-1) = \frac{(n-1)(n-2)}{2} \;\; .
\eqno(3.2)$$

As you see for the first time one observable phase arrives in the
case of 3 quark-lepton generations.

Now a bit of history. At the time when Cabibbo used the mixing
angle $\theta_c$ \cite{1} only three quarks ($u$, $d$ and $s$)
were known, and his suggestion was to mix $d$- and $s$-quarks in
the expression for the charged quark current: $$J_\mu^+ = \bar u
\gamma_\mu(1+\gamma_5)[d \cos\theta_c + s\sin\theta_c] \;\; .
\eqno(3.3) $$

In this way he related the suppression of the strange particles
weak interactions to the smallness of angle $\theta_c$,
$\sin^2\theta_c \approx 0.05$. After the establishment of
Glashow-Salam-Weinberg theory of the weak interactions through the
weak neutral current discovery in 1973 and discovery of a charm
quark in 1974 it became clear, that 2 quark-lepton generations
exist. From the point of view of GSW theory the mixing of quark
generations should be described by the unitary 2$\times$2 matrix
which according to eqs. (3.1, 3.2) has one angle and zero
observable phases. This angle is Cabbibo angle. However even
before the $c$-quark discovery Kobayashi and Maskawa noticed that
in order to describe CP-violation (CPV) Standard Model needs at
least 3 quark-lepton generations since for the first time the
observable phase shows up for $n=3$ \cite{2}. At that time CPV was
known only in $K^0$-mesons and to test KM mechanism one needed
other systems. Finally almost 30 years after KM model of CP
violation was suggested it was confirmed by the magnitude of CPV
in neutral $B$-mesons.

In the case of three generations the matrix of charged currents
looks like:

$$ \overline{(u c t)_L} \left(\begin{array}{lll} V_{ud} & V_{us} &
V_{ub}
\\ V_{cd} & V_{cs} & V_{cb} \\ V_{td} & V_{ts} & V_{tb}
\end{array} \right)
 \left( \begin{array}{c} d \\ s \\ b \end{array} \right)_L \;\; ,
 \eqno(3.4)
 $$
where the matrix elements $V_{ik}$ depend on four parameters:
three angles and one phase. Let us present one possible way of the
parametrization of matrix $V$, which is called ``standard''
parametrization (straightforwardly generalizable for $n>3$
\cite{4}). It is achieved by consequent rotations in planes (12),
(13) and (23) and the rotation in plane (13) is accomplished by
the phase rotation. Performed in such a way the phase rotation
cannot be removed by $U(1)$ rotations of the quark fields:
 $$
 V = R_{23} \times R_{13} \times R_{12} \;\; , \eqno(3.5)
 $$
 $$
 R_{23} = \left(\begin{array}{ccc} 1 & 0 & 0 \\ 0 & c_{23} &
 s_{23} \\ 0 & -s_{23} & c_{23} \end{array} \right) \;\; ,
$$
$$ R_{13} =
 \left( \begin{array}{ccc} c_{13} & 0 & s_{13} e^{-i\delta} \\
 0 & 1 & 0 \\
 -s_{13}e^{i\delta} & 0 & c_{13} \end{array} \right) \;\; ,
 \;\;
 R_{12} = \left(\begin{array}{ccc} c_{12} & s_{12} & 0 \\
  -s_{12} & c_{12} & 0 \\
  0 & 0 & 1 \end{array} \right) \;\; , \eqno(3.6)
  $$
and, finally:
 $$ V = \left( \begin{array}{ccc} c_{13} c_{12} &
c_{13} s_{12} & s_{13} e^{-i\delta} \\ -c_{23} s_{12} -s_{23}
s_{13} c_{12} e^{i\delta} & c_{23} c_{12} -s_{12} s_{13} s_{23}
e^{i\delta} & s_{23} c_{13} \\ s_{12} s_{23} -c_{12} c_{23} s_{13}
e^{i\delta} & -s_{23} c_{12} -c_{23} s_{13} s_{12} e^{i\delta} &
c_{23} c_{13}
\end{array} \right) \; . \eqno(3.7)
$$

As the next step we can take the experimental values for $V_{ik}$
and extract three angles $\theta_{12}$, $\theta_{23}$,
$\theta_{13}$ and phase $\delta$ from them. The system is
overconstrained; we have more than 4 experimental numbers (see
below) and in this way one can check how good CKM model in
describing data is. However it appeared to be useful to
reparametrize $V_{ik}$ with the help of the so-called Wolfenstein
parametrization. Let us introduce new parameters $\lambda$, $A$,
$\rho$ and $\eta$ according to the following definitions: $$
\lambda \equiv s_{12} \; , \;\; A\equiv \frac{s_{23}}{s_{12}^2}
\;\; , \;\; \rho = \frac{s_{13}}{s_{12}s_{23}}\cos\delta \;\; , $$
$$ \eta = \frac{s_{13}}{s_{12}s_{23}}\sin\delta \;\; , \eqno(3.8)
$$ and get expressions for $V_{ik}$ through $\lambda$, $A$, $\rho$
and $\eta$: $$ V = \left(\begin{array}{lll} V_{ud} & V_{us} &
V_{ub}
\\ V_{cd} & V_{cs} & V_{cb} \\ V_{td} & V_{ts} & V_{tb}
\end{array} \right) \approx \left( \begin{array}{ccc}
1-\lambda^2/2 & \lambda & A\lambda^3(\bar\rho -i\bar\eta) \\
-\lambda -iA^2 \lambda^5 \bar\eta & 1-\lambda^2/2 & A\lambda^2 \\
A\lambda^3(1-\bar\rho -i\bar\eta) & -A\lambda^2 -iA\lambda^4
\bar\eta & 1 \end{array} \right) \;\; , \eqno(3.9) $$ where
$$\bar\rho \equiv \rho(1-\frac{\lambda^2}{2}) \;\;, \; \bar\eta
\equiv \eta(1-\frac{\lambda^2}{2})\;\; . \eqno(3.10) $$

Obtaining the last expression we take into account the following
hierarchy of angles $\theta_{ij}$: $s_{13} \ll s_{23} \ll s_{12}
\ll 1$ (see below). This last form of CKM matrix is very
convenient for qualitative estimates and numerically  is rather
accurate.

The unitarity of the matrix $V$ leads to the following six
equations that can be drawn as triangles on a complex plane (under
each term in these equations the power of $\lambda$ entering it,
is shown):

$$
\begin{array}{cccccc}
 V_{ud}^*
V_{us} & + & V_{cd}^* V_{cs} & + & V_{td}^* V_{ts} & = 0 \;\; , \\
 \sim\lambda & & \sim\lambda &
& \sim\lambda^5 &
\end{array}
\eqno(3.11) $$ $$
\begin{array}{cccccc}
V_{ud}^* V_{ub} & + & V_{cd}^* V_{cb} & +
& V_{td}^* V_{tb} & = 0 \;\; , \\
\sim\lambda^3 & & \sim\lambda^3 & & \sim\lambda^3
\end{array}
\eqno(3.12) $$ $$
\begin{array}{cccccc}
V_{us}^* V_{ub} & + & V_{cs}^* V_{cb} & + & V_{ts}^* V_{tb} & =
0 \;\; ,\\ \sim\lambda^4 & & \sim\lambda^2 & &
\sim\lambda^2 &
\end{array}
\eqno(3.13) $$ $$
\begin{array}{cccccc}
V_{ud} V_{cd}^* & + & V_{us} V_{cs}^* & + & V_{ub} V_{cb}^* & = 0 \;\; , \\
\sim\lambda & & \sim\lambda & &
\sim\lambda^5 &
\end{array}
\eqno(3.14) $$ $$
\begin{array}{cccccc}
V_{ud} V_{td}^* & + & V_{us} V_{ts}^*
& + & V_{ub} V_{tb}^* & = 0 \;\; , \\
\sim\lambda^3 & & \sim\lambda^3 & & \sim\lambda^3 &
\end{array}
\eqno(3.15) $$ $$
\begin{array}{cccccc}
V_{cd} V_{td}^* & + & V_{cs} V_{ts}^* & + & V_{cb} V_{tb}^* & = 0
\;\; . \\
\sim\lambda^4 & & \sim\lambda^2 & &
\sim\lambda^2 &
\end{array}
\eqno(3.16) $$

Among these triangles the four are almost degenerate: one side is
much shorter than two others, and two triangles, expressed by
equations (3.12) and (3.15), have all three sides of more or less
equal lengths, of the order of $\lambda^3$. These two
nondegenerate triangles almost coincide. To prove this statement
let us note that with very good accuracy $V_{ud} = V_{tb}=1$, that
is why two sides of these triangles have equal lengths and
directions. That is why the triangles coincide and their third
sides  should also be equal: $$V_{cd}^* V_{cb} = V_{us}V_{ts}^*
\;\; . \eqno(3.17) $$

Now, since $V_{us}$ and $-V_{cd}$ are almost equal to each other
(and to $\lambda$) we come to the following result: $$ V_{ts}^* =
-V_{cb} \;\; , \eqno(3.18) $$ the validity of which can be checked
by eq.(3.9).

So, as a result we have only one nondegenerate unitarity triangle;
it is usually described by a complex conjugate of our equation
(3.12): $$V_{ud}V_{ub}^* +V_{cd} V_{cb}^* +V_{td}V_{tb}^* = 0 \;\;
, \eqno(3.19) $$ and it is shown in Fig.2. It has the angles which
are called $\beta$, $\alpha$ and $\gamma$ (according to BaBar
collaboration) or $\phi_1$, $\phi_2$ and $\phi_3$ (according to
Belle collaboration). They are determined from CPV asymmetries in
$B$-mesons decays.

\begin{figure}[!htb]
\centering
\epsfig{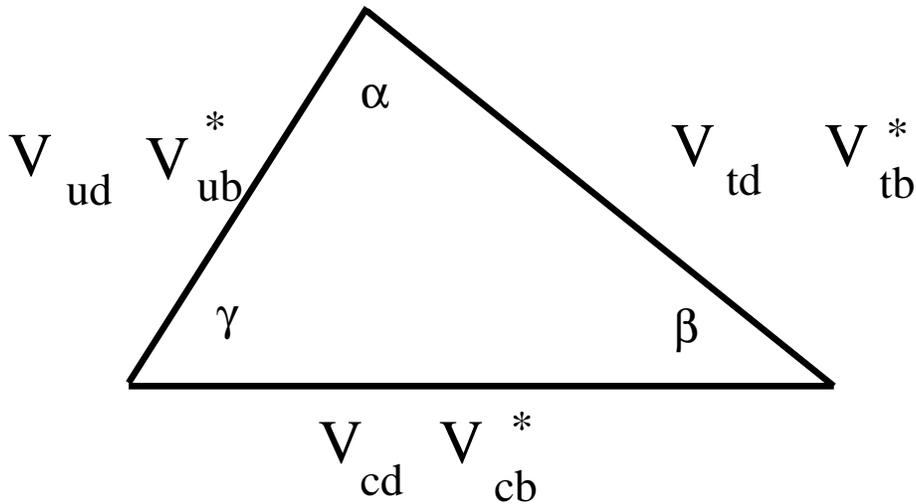}
\caption{\em Unitarity triangle }
\label{WW2Fermi}
\end{figure}

 Looking at Figure 2 one can easily obtain the
following formulas:

$$ \beta = \pi - \arg\frac{V_{tb}^* V_{td}}{V_{cb}^* V_{cd}} =
\phi_1 \;\; , \eqno(3.20) $$

$$\alpha = \arg\frac{V_{tb}^* V_{td}}{-V_{ub}^* V_{ud}} = \phi_2
\;\; , \eqno(3.21) $$

$$\gamma = \arg\frac{V_{ub}^* V_{ud}}{-V_{cb}^* V_{cd}} = \phi_3
\;\; . \eqno(3.22) $$

Angle $\beta$ ($\phi_1$) was directly measured through time
dependent CPV asymmetry in $B_d \to J/\Psi K$ decays, $\alpha$
($\phi_2$) has been measured recently with a rather poor accuracy
from CPV asymmetries in $B_d \to\pi^+ \pi^-$ decays and, finally,
$B_s$ decays could be important to determine angle $\gamma$
($\phi_3$).

Let us make three final remarks about the unitarity triangle:
\begin{enumerate}
\item In standard parametrization which we use $V_{cd}V_{cb}^*$ is
almost real;
\item Multiplication of any up quark field on a phase does not
change the unitarity triangle while multiplying  $d$- or $b$-quark
field on a phase we rotate it as a whole not changing its angles
which are physical observables;
\item Usually a rescaled triangle is used. We get it dividing all
three sides by $|V_{cb}^* V_{cd}| \approx A\lambda^3$. In this way
the length of the triangle basement becomes equal to one while two
other sides have the length of the order of one.
\end{enumerate}

Four quantities are needed to specify CKM matrix: $s_{12}, s_{13},
s_{23}$ and $\delta$, or $\lambda, A, \rho, \eta$. Knowing more we
are checking Standard Model and looking for New Physics.

\section{$\mbox{\boldmath$V_{us}, V_{cb}, V_{ub}$}$ --
first circle}

The most precise value for the quantity $V_{us}$ follows from the
extrapolation of the formfactor of $K \to \pi e\nu$ decay
$f_+(q^2)$ to the point $q^2 =0$, where $q$ is the lepton pair
momentum. Due to the Ademollo-Gatto theorem corrections to the CVC
value $f_+(0)=1$ are of the second order of flavor SU(3)
violation, and these small terms were calculated in \cite{5}. As a
result of this analysis Particle Data Group gives the following
value \cite{6}: $$V_{us} \equiv \lambda = 0.2196 \pm 0.0026 \;\; .
\eqno(4.1) $$

The accuracy of $\lambda$ is high: the other parameters of CKM
matrix are known much worse.

The value of $V_{cb}$ is determined from the inclusive and
exclusive semileptonic decays of $B$-mesons with charmed particles
production. At the level of quarks $b\to ce\nu$ transition is
responsible for these decays.

According to PDG \cite{6}: $$V_{cb} = (41.2 \pm 2.0) 10^{-3} \;\;
; \eqno(4.2) $$ and the error is dominated by a theoretical one.

From eq.(3.9) with the help of (4.2) for parameter $A$ we get: $$
A = \frac{V_{cb}}{\lambda^2} = 0.85 \pm 0.04 \;\; . \eqno(4.3) $$

From the formula for the semileptonic width: $$ \Gamma_{SL} \sim
|V_{cb}|^2 (m_b - m_c)^5 \eqno(4.4) $$ it is clearly seen that 4\%
error in $V_{cb}$ corresponds to the knowledge of quark mass
difference with the very high accuracy: $$m_b - m_c = 3 \; {\rm
GeV} \pm 60 \; {\rm MeV} \;\; . \eqno(4.5) $$

Vast literature is devoted to the different aspects of the
$V_{cb}$ determination. Not going into details let us only note
that at the moment the error in $V_{cb}$ is not a ``bottle neck''
since the errors in other relevant quantities ($V_{ub}$, $V_{td}$,
$\sin 2\beta$) are significantly larger.

The value of $|V_{ub}|$ is extracted from the semileptonic
$B$-mesons decays without the charmed particles in the final state
which originated from $b\to ul\nu$ transition. $b\to cl\nu$ decays
are approximately 100 times more probable than $b\to ul\nu$ and
for their suppression the high energy charged lepton tail is
examined (the energetic leptons cannot accompany heavy
$D$-mesons). Theoretical analysis of such  semiinclusive decays is
highly involved and it leads to a large theoretical uncertainty
\cite{6}: $$\mid\frac{V_{ub}}{\lambda V_{cb}}\mid = 0.40 \pm 0.08
\;\; . \eqno(4.6) $$

There are hopes to considerably diminish this error using the
exclusive modes $B\to \pi e\nu$ and $B\to \omega e\nu$. (The
recent study of the $B\to \pi e\nu$ and $B\to \rho e\nu$ decays by
CLEO collaboration gives 10\% smaller central value of $V_{ub}$
with practically the same error \cite{6'}.

Using equation (3.9) we obtain a bound on the parameters
$\bar\rho$ and $\bar\eta$: $$ \sqrt{\bar\rho^2 +\bar\eta^2} = 0.40
\pm 0.08 \;\; , \eqno(4.7) $$ which produces a circle on
$(\bar\rho, \bar\eta)$ plane with the center at the point $(0,0)$.
The area between  such two circles  in Fig.1 corresponds to the
domain allowed at one sigma.

The detailed articles on $V_{ub}$ and $V_{cb}$ determination with
bibliography can be found in \cite{6}; for the recent review see
\cite{5'}.

To finish this section let us look at the value of $\mid
V_{ud}\mid$. Its value extracted from the neutron decays has the
smallest theoretical uncertainty \cite{6''}: $$ \mid V_{ud}\mid =
0.9713(13) \;\; ; \eqno(4.8) $$ using this value as well as (4.1),
(4.2) and (4.6) we obtain: $$  \mid V_{ud}\mid^2 +\mid
V_{us}\mid^2 +\mid V_{ub}\mid^2  =  0.9434(25) +0.0482(11)
+0.00001 = $$ $$  =  0.9916(27) \;\; , \eqno(4.9)  $$ and the
unitarity of CKM matrix is violated at $3\sigma$ level. This can
be a signal for New Physics, while it may be a statistical
fluctuation as well. PDG discussing the value of $V_{ud}$ takes
into account the nuclear beta decays as well resulting in
\cite{6}: $$ \mid V_{ud}\mid = 0.9734(8) \;\; , \eqno(4.10) $$
which diminishes the violation of unitarity to $2\sigma$.

\section{CPV: history; why phases are relevant}

In 1956 Lee and Yang in order to solve $\theta-\tau$ problem
suggested that P-parity is broken in weak interactions
\cite{6'''}. Soon it was noted that C-invariance is also broken in
a proposed theory \cite{6''''}.

In 1957 looking for a way to resurrect P-invariance L.D.~Landau
stated that weak interactions should be invariant under the
product of P reflection and C conjugation. He called this product
the combined inversion and according to him it should substitute
$P$-inversion broken in weak interactions \cite{7}. In this way
the theory should be invariant when together with changing sign of
the coordinate, $\bar r \to -\bar r$, one changes an electron to
positron, proton to antiproton and so on.

It is clearly seen from paper \cite{7} that according to Landau
CP-invariance should become a basic symmetry for physics in
general and weak interactions in particular.

This beautiful picture did not stop experimentalists (or may be
even stimulated) and in 1964 CP violating decay of the long-lived
neutral kaon on two pions was discovered \cite{8}. The question of
the violation of P-parity in weak interactions which stimulated
Landau was: why is P violated? Landau's answer to the question
``Why is parity violated in weak interactions'' was: because CP,
not P is the fundamental symmetry of nature. A modern answer to
the same question is: because in P-invariant theory with the Dirac
fermions the gauge invariant mass terms can be written for quarks
and leptons which are not protected of being of the order of
$M_{\rm GUT}$ or $M_{\rm Planck}$. So in order to have our world
made from light particles P-parity should be violated.

$K_L \to 2\pi$ decay discovered in 1964 occurs due to CPV in the
mixing of neutral kaons ($\tilde\varepsilon \neq 0$). Only thirty
years later the second major step was done: direct CPV was
observed in kaon decays \cite{9}: $$\frac{\Gamma(K_L \to \pi^+
\pi^-)}{\Gamma(K_S \to \pi^+ \pi^-)} \neq \frac{\Gamma(K_L \to
\pi^0 \pi^0)}{\Gamma(K_S \to \pi^0 \pi^0)} \; , \;\;
\varepsilon^\prime \neq 0 \eqno(5.1) $$

Finally, in the year 2001 CPV was for the first time observed out
of the decays of neutral kaons: the time dependent CP-violating
asymmetry in $B^0$ decays was measured \cite{9'}:
$$a(t)=\frac{N(B^0 \to J/\Psi K_{S(L)})-N(\bar B^0 \to J/\Psi
K_{S(L)})}{N(B^0 \to J/\Psi K_{S(L)})+ N(\bar B^0 \to J/\Psi
K_{S(L)})}\neq 0 \;\; . \eqno(5.2)$$

Starting from the year 1964 we know that there is no symmetry
between particles and antiparticles. In particular, the
$C$-conjugated partial widths are different: $$\Gamma(A \to BC)
\neq \Gamma(\bar A \to \bar B \bar C) \;\; . \eqno(5.3) $$

However CPT (deduced from the invariance of the theory under
4-dimen\-sional rotations) remains intact. That is why the total
widths as well as the masses of particles and antiparticles are
equal: $$M_A = M_{\bar A} \;, \;\; \Gamma_A = \Gamma_{\bar A} ~~~~
({\rm CPT}) \;\; . \eqno(5.4) $$

The consequences of CPV can be divided into macroscopic and
microscopic. CPV is one of the three famous Sakharov's necessary
conditions to get as a result of evolution of charge symmetric
Universe a charge nonsymmetric one \cite{9''}. In these lectures
we will not discuss this very interesting and well developed
branch of physics, but will deal with CPV in particle physics
where the data obtained up to now confirm Kobayashi-Maskawa model
of CPV. New data which should become available in coming years
could well disprove it clearly demonstrating the necessity of
physics beyond the Standard Model. Let us note that there exist
many excellent reviews on CPV; a partial list of them can be found
in ref. \cite{10}, which starts from the lectures on this topic
given at ITEP Winter Schools in a chronological order.

The next question we wish to discuss is why the phases are
relevant for CPV. Let us take the simplest example of one scalar
particle decaying in two scalar particles. The interaction
Lagrangian is: $$ {\cal L} = \lambda AB^* C^* + \lambda^* A^* B C
\;\; . \eqno(5.5) $$

Performing CP conjugation (changing particles to antiparticles and
reflecting space coordinates) we obtain: $${\cal L}_{\rm CP} =
\lambda A^* B C + \lambda^* AB^* C^* \;\; , \eqno(5.6) $$ which
coincides with the original Lagrangian only if $\lambda$ is real,
$\lambda^* = \lambda$. That is why the complex coupling constants
are necessary for CPV. However, this complexity is not enough: in
our example we can redefine field $A$ by phase rotation, $A =
e^{i\psi}A^\prime$, in this way making $\lambda$ real. However,
adding one new field (say $D$) we introduce many new couplings
($\lambda_1 DBC + \lambda_2 DAB + \lambda_3 DAC$) while by
rotation  $D = e^{i\alpha} D^\prime$ a phase of only one coupling
constant can be eliminated. Thus to get CPV several fields are
needed. The last statement is illustrated in Standard Model by the
fact that at least three generations are needed to get CPV through
the phases in the quark mixing matrix.

\section{$\mbox{\boldmath$M^0 - \bar M^0$}$ mixing; CPV in
mixing}

In this part the general formulas for the meson-antimeson mixing
will be derived. In order to mix the mesons must be neutral, not
coincide with its antiparticle and decay due to the weak
interactions. There are four such pairs: $$ K^0(\bar s d) - \bar
K^0(s\bar d) \; , \;\; D^0(c\bar u) - \bar D^0(\bar c u) \;\; , $$
$$ B_d^0(\bar b d) - \bar B_d^0(b\bar d) \;\;  {\rm and} \;\;
B_s^0(\bar b s) - \bar B_s^0(b \bar s) \;\; . $$

Fast $t\to bW$ decay prevents forming $t$-quark containing
hadrons.

Mixing occurs in the second order in weak interactions through the
box diagram which is shown in Fig. 3 for $K^0 - \bar K^0$ pair.

\begin{figure}[!htb]
\centering
\epsfig{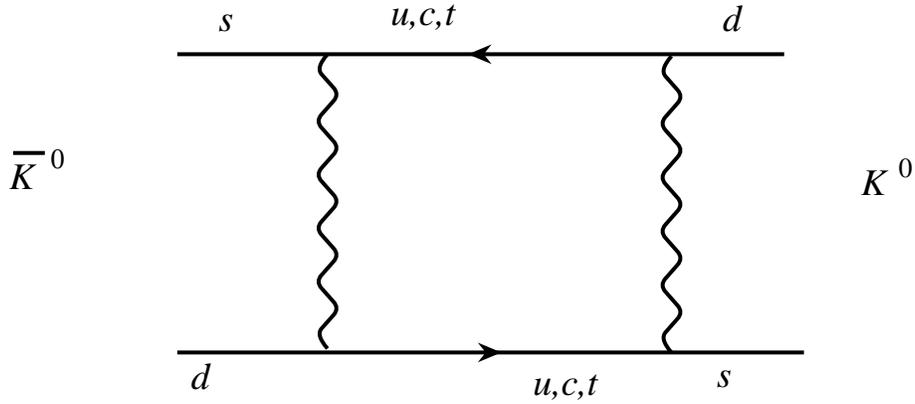}
\caption{\em $K^0 - \bar K^0$ mixing }
\label{WWFermi}
\end{figure}

The effective $2\times 2$ hamiltonian $H$ is used to describe the
meson-antimeson mixing. It is most easily written in the following
basis: $M^0 = \left(\begin{array}{c} 1 \\ 0
\end{array} \right)$, $\bar M^0 = \left(\begin{array}{c} 0 \\ 1
\end{array} \right)$. The meson-antimeson system evolves according
to the Shroedinger equation with this effective hamiltonian which
is not hermitian since it takes meson decays into account. So, $H=
M -\frac{i}{2}\Gamma$,  where both $M$ and $\Gamma$ are hermitian.

According to CPT invariance the diagonal elements of $H$ are
equal: $$< M^0 \mid H\mid M^0 > = <\bar M^0 \mid H \mid \bar M^0
> \;\; . \eqno(6.1) $$

Substituting into the Shroedinger equation
$$i\frac{\partial\psi}{\partial t} = H\psi \eqno(6.2) $$ $\psi$ --
function in the following form: $$\psi = \left(\begin{array}{c} p
\\ q \end{array}\right) e^{-i\lambda t} \eqno(6.3) $$
we come to the following equation:

$$ \left( \begin{array}{cc}M - \frac{i}{2} \Gamma & M_{12}
-\frac{i}{2} \Gamma_{12} \\ ~ &  ~\\ M_{12}^* - \frac{i}{2}
\Gamma_{12}^* & M -\frac{i}{2} \Gamma  \end{array} \right) \left(
\begin{array}{c} p \\  ~  \\ q \end{array} \right) = \lambda \left(
\begin{array}{c} p \\ ~  \\ q \end{array} \right) \eqno(6.4) $$
from which for eigenvalues ($\lambda_\pm$) and eigenvectors
($M_\pm$) we obtain: $$ \lambda_{\pm} = M -\frac{i}{2}\Gamma \pm
\sqrt{(M_{12} - \frac{i}{2}\Gamma_{12})(M_{12}^* -
\frac{i}{2}\Gamma_{12}^*)} \; ,  \eqno(6.5)$$ $$ \left\{
\begin{array}{l} M_+ = pM^0 + q \bar M^0 \\ M_- = pM^0 - q \bar
M^0 \end{array} \right. \;\; , \;\; \frac{q}{p} =
\sqrt{\frac{M_{12}^* - \frac{i}{2}\Gamma_{12}^*}{M_{12}
-\frac{i}{2}\Gamma_{12}}} \;\; . \eqno(6.6) $$

Multiplying $M^0$ by $e^{i\phi}$ we are changing the phase of the
ratio $q/p$, that is why $\arg\left(\frac{q}{p}\right)$ is not a
physical observable.

CP transforms field $M^0$ in the following way: $$CPM^0 =
e^{i\alpha} \bar M^0 \;\; , \eqno(6.7) $$ being defined up to the
arbitrary phase $\alpha$. We can rotate field $M^0$ by phase
$\alpha/2$, removing phase $\alpha$ from the definition of CP
transformation. In this way we come to the standard definition of
CP transformation: $$CPM^0 = \bar M^0 \;\; , \eqno(6.8) $$
simultaneously loosing freedom of the arbitrary phase rotation of
field $M^0$.

The eigenstates $M_+$ and $M_-$ are not orthogonal in general
case: $$< M_+ \mid M_-
> = \mid p\mid^2 -\mid q\mid^2 \neq 0 \;\; . \eqno(6.9) $$ However
if there is no CPV in mixing, then: $$ <M^0\mid H \mid \bar M^0 >
= < \bar M^0 \mid H \mid M^0 >  \;\; , \eqno(6.10) $$ $$ M_{12} -
\frac{i}{2} \Gamma_{12} = M_{12}^* - \frac{i}{2} \Gamma_{12}^*
\;\; , \eqno(6.11) $$ that is why: $$ \mid\frac{p}{q}\mid = 1
\;\; , \;\; <M_+ \mid M_-
> = 0 \;\; . \eqno(6.12) $$

We observe the one-to-one correspondence between CPV in mixing and
nonorthogonality of the eigenstates $M_+$ and $M_-$. According to
Quantum Mechanics if two hermitian matrices $M$ and $\Gamma$
commute, they have a common orthonormal basis. Let us calculate
the commutator of $M$ and $\Gamma$: $$ [M, \Gamma] =\left(
\begin{array}{cc} M_{12} \Gamma_{12}^* - M_{12}^* \Gamma_{12} & 0
\\ ~ \\ 0 & M_{12}^* \Gamma_{12} - M_{12} \Gamma_{12}^*
\end{array}\right) \; . \eqno(6.13) $$
It equals zero if the phases of $M_{12}$ and $\Gamma_{12}$
coincide modulo $\pi$. So, for $[M \Gamma] =0$ we get $\mid
q/p\mid =1$, $<M_+ \mid M_- > =0$ and there is no CPV in the
meson-antimeson mixing. And vice versa.

Introducing quantity $\tilde\varepsilon$ according to the
following definition: $$\frac{q}{p} =
\frac{1-\tilde\varepsilon}{1+\tilde\varepsilon} \;\; , \eqno(6.14)
$$ we see that if $Re~\tilde\varepsilon \neq 0$, then CP is
violated. From (6.6) for the eigenstates we obtain: $$ M_+ =
\frac{1}{\sqrt{1+ \mid \tilde\varepsilon \mid^2}} \left[\frac{M^0
+\bar M^0}{\sqrt{2}} + \tilde\varepsilon \frac{M^0 - \bar
M^0}{\sqrt{2}} \right] \;\; ,  $$ $$ M_- = \frac{1}{\sqrt{1+ \mid
\tilde\varepsilon \mid^2}} \left[\frac{M^0 -\bar M^0}{\sqrt{2}} +
\tilde\varepsilon \frac{M^0 +\bar M^0}{\sqrt{2}} \right] \;\; .
\eqno(6.15) $$

If CP is conserved, then $Re~\tilde\varepsilon = 0$, $M_+$ is CP
even and $M_-$ is CP odd (CP transformation should be defined
according to (6.7) - (6.8)). If CP is violated in mixing, then
$Re~\tilde\varepsilon \neq 0$ and $M_+$ and $M_-$ get admixtures
of the opposite CP parity and become nonorthogonal.

\section{Space-time pattern of $\mbox{\boldmath$K^0 - \bar
K^0$}$ \\ ($\mbox{\boldmath$B^0 - \bar B^0 \; , \;\; \nu_\mu -
\nu_e$}$) oscillations}

Neglecting CPV for the states with the definite masses and width
we obtain: $$ \left\{ \begin{array}{l} K_1 =
\frac{1}{\sqrt{2}}(K^0 + \bar K^0) \;\; , \\ K_2 =
\frac{1}{\sqrt{2}}(K^0 - \bar K^0) \;\; ,
\end{array} \right. \eqno(7.1)
$$ where CP$|K^0> = |\bar K^0>$, CP$|\bar K^0> = |K^0>$.

A plain wave which describes the propagation of the superposition
of $K_1$ and $K_2$ with definite energies and momenta is: $$
\Psi(x,t) = e^{-iE_1 t +i p_1 x} \mid K_1 > + e^{-i E_2 t + i p_2
x} \mid K_2
> \;\; , \eqno(7.2) $$
$$ E_1^2 - p_1^2 = m_1^2 \; , \;\; E_2^2 - p_2^2 = m_2^2 \;\; , $$
and we consider the propagation in the direction of axis $x$. We
should impose a boundary condition stating that at point $x=0$
continuously in time $K^0$ is produced (or $B^0$, or $\nu_e$). The
only way to get this is by putting $E_1 = E_2$, having at $x=0$:
$$ e^{-iEt}\mid K_1 > + e^{-iEt}\mid K_2 > = e^{-iEt}[\mid K_1 > +
\mid K_2 >] = e^{-iEt}\mid K^0
> \;\; . \eqno(7.3) $$

If we do not impose this boundary condition and take $E_1 \neq
E_2$, then after time $\tau \sim \frac{1}{E_1 - E_2}$ at the point
$x=0$ $\bar K^0$ will emerge, while in the experiment only the
events when $K^0$ at $x=0$ is produced are selected. For example,
in CPLEAR we have: $p\bar p \to \pi K^- K^0$ and $K^0$ is tagged
by $K^-$, showing that $K^0$ was produced in the interaction
point, and not $\bar K^0$. Another example: in the reaction $p
n\to p\Lambda K^0$ together with $\Lambda$ hyperon $K^0$ is
produced and not $\bar K^0$ because of strangeness conservation in
strong interactions. In the case of antineutrino production in
nuclear reactor initiated by $n\to p e^- \bar\nu_e$ transition it
is always electron antineutrino and never muon antineutrino.

Substituting $E_1 = E_2 = E$ in equation (7.2) we get: $$
\Psi(x,t) = e^{-iEt + ip_1 x}[\mid K_1 > +e^{i(p_2 -p_1)x}\mid K_2
>] \;\; . \eqno(7.4)
$$

For the phase factor in brackets we have: $$ p_2 -p_1 = \sqrt{E^2
-m_2^2} - \sqrt{E^2 -m_1^2} =$$ $$ = \sqrt{E^2 -m_1^2 + m_1^2
-m_2^2} - \sqrt{E^2 -m_1^2} = \frac{m_1^2 - m_2^2}{2 p} \;\; ,
\eqno(7.5) $$ where we take into account that for mesons (as well
as for neutrinos) always $E^2 - m_1^2 \gg m_1^2 - m_2^2$.
Therefore for a probability to detect $K^0$ at a distance $x$ from
a production point we have: $$ P_{K^0 K^0} = \frac{1}{2}
\left[1+\cos(\frac{m_1^2 - m_2^2}{2p}x)\right]\left
|_{_{_{_{\large \Delta m \ll m}}}}  =
\frac{1}{2}\left[1+\cos(\frac{\Delta m}{\beta\gamma}x)\right]
\right. \;\; , \eqno(7.6) $$ $$ \beta = v/c \; , \;\; \gamma =
1/\sqrt{1-\beta^2} \;\; , $$ where the second equality in eq.(7.6)
holds for mesons, but (may be) not for neutrinos.

There is no surprise that both neutrino and meson oscillations are
described by the identical formulas. In the first paper where the
neutrino oscillations were considered \cite{11}, B.M.~Pontecorvo
did it in full analogy with $K^0 - \bar K^0$ oscillation analysis
of Gell-Mann and Pais. Since at that time the second (muon)
neutrino was not yet discovered Pontecorvo considered $\nu_e -
\bar\nu_e$ oscillations which were allowed because $V-A$ theory
was not established yet (in $V-A$ theory $\nu - \bar\nu$
oscillations are forbidden by chirality conservation). In this
particular case the diagonal elements of $\nu -\bar\nu$ mixing
matrix should be equal due to CPT and mixing is maximal,
$\theta_\nu = \pi/4$, just as for $K^0(B_d, B_s)$ mesons.

\section{$\mbox{\boldmath$K^0 - \bar K^0$}$ mixing, $\mbox{\boldmath$\Delta
m_{LS}$}$}

$\Gamma_{12}$ for the $K^0 - \bar K^0$ system is given by the
diagram shown in Fig. 4. With our choice of CKM matrix $V_{us}$
and $V_{ud}$ are real, so $\Gamma_{12}$ is real.

\begin{figure}[!htb]
\centering
\epsfig{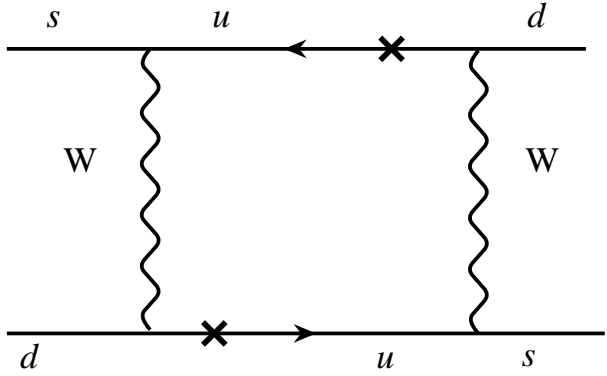}
\caption{\em Quark diagram responsible for $\Gamma_{12}$ in
$K^0 - \bar K^0$ system }
\label{WW4Fermi}
\end{figure}

$M_{12}$ is given by dispersion part of the  diagram shown in
Fig.5. Now all three up quarks should be taken into account.

\begin{figure}[!htb]
\centering
\epsfig{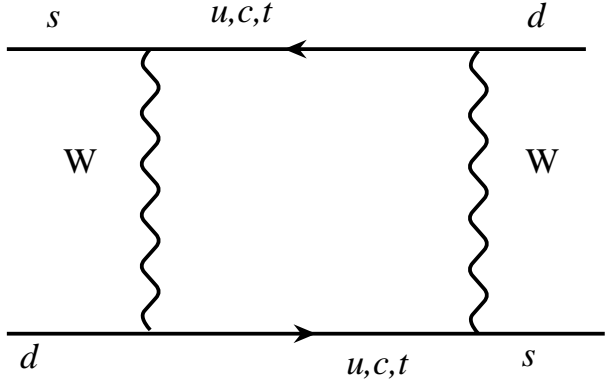}
\caption{\em Quark diagram responsible for $M_{12}$ in
$K^0 - \bar K^0$ system }
\label{WW5Fermi}
\end{figure}

 To calculate this
diagram it is convenient to implement Glashow-Illiopulos-Maiani
compensation mechanism from the very beginning, subtracting zero
from the sum of the fermion propagators: $$ \frac{V_{us}
V_{ud}^*}{\hat p - m_u} + \frac{V_{cs} V_{cd}^*}{\hat p - m_c} +
\frac{V_{ts} V_{td}^*}{\hat p - m_t} - \frac{\sum\limits_i V_{is}
V_{id}^*}{\hat p} \;\; . \eqno(8.1) $$

Since $u$-quark is with good accuracy massless, $m_u \approx 0$,
its propagator drops out and we are left with the modified $c$-
and $t$-quark propagators: $$ \frac{1}{\hat p - m_{c,t}}
\longrightarrow \frac{m_{c,t}^2}{(p^2 - m_{c,t}^2) \hat p} \;\; .
\eqno(8.2) $$

The modified fermion propagators decrease in ultraviolet so
rapidly that one can calculate the box diagrams in the unitary
gauge, where $W$-boson propagator does not decrease (one can
demonstrate that the diagrams with the charged higgs exchange
which occur in the renormalizable $R$ gauges, and for which
eq.(8.1) does not work since the vertex of higgs emission is
proportional to the quark masses, becomes zero in the limit $\xi
\to \infty$, which corresponds to the unitary gauge).

We easily get the following estimates for three remaining diagram
contributions in $M_{12}$: $$ (cc): \;\; \lambda^2(1+2i\bar\eta
A^2 \lambda^4) G_F^2 m_c^2 \;\; , \eqno(8.3) $$ $$ (ct): \;\;
\lambda^6(1-\bar\rho -i\bar\eta) G_F^2 m_c^2
\ln(\frac{m_t}{m_c})^2 \;\; , \eqno(8.4) $$ $$ (tt): \;\;
\lambda^{10}(1-\bar\rho -i\bar\eta)^2
 G_F^2 m_t^2 \;\; . \eqno(8.5) $$

Since $m_c \approx 1.3$ GeV and $m_t \approx 175$ GeV we observe
that the $cc$ diagram dominates in $Re M_{12}$ while $Im M_{12}$
is dominated by ($tt$) diagram. The real part dominates in
$M_{12}$: $$ \frac{Im M_{12}}{Re M_{12}} \sim \lambda^8
\left(\frac{m_t}{m_c}\right)^2\sim 0.1 \;\; . \eqno(8.6) $$

The explicit calculation of the $cc$ exchange diagram gives: $$
{\cal L}_{\Delta s =2}^{\rm eff} = -\frac{g^4}{2^9 \pi^2
M_W^4}(\bar s \gamma_\alpha(1+\gamma_s)d)^2 \eta_1 m_c^2 V_{cs}^2
V_{cd}^{*^2} \;\; , \eqno(8.7) $$ where $g$ is SU(2) gauge
coupling constant, $g^2/8 M_W^2 = G_F/\sqrt 2$, and the factor
$\eta_1$ takes into account the hard gluon exchanges. Since $$
M_{12}-\frac{i}{2}\Gamma_{12} = <K^0\mid H^{eff}\mid \bar K^0 >
/2m_K\footnote{Factor $1/2 m_K$ appears when taking a square root
from the quadratic in meson masses mesonic Hamiltonian in order to
obtain linear in meson masses Hamiltonian which enters the
Shroedinger equation: $$\left[
\begin{array}{cc} (M-\frac{i}{2}\Gamma)^2 & <K^0\mid H^{eff}\mid\bar
K^0 > \\ <\bar K^0\mid H^{eff}\mid K^0> & (M-\frac{i}{2}\Gamma)^2
\end{array}\right]^{1/2} = \left( \begin{array}{cc}
M-\frac{i}{2}\Gamma & \frac{<K^0\mid H^{eff}\mid\bar K^0 >}{2m_K}
\\ \frac{<\bar K^0\mid H^{eff}\mid K^0>}{2m_K} &
M-\frac{i}{2}\Gamma
\end{array} \right) \;\; , $$ here $H^{eff} = -{\cal L}_{\Delta S
=2}^{eff}$.} \;\; , \eqno(8.8) $$ (here $H^{eff} = -{\cal
L}_{\Delta s =2}^{eff}$) we should calculate the matrix element of
the product of two $V-A$ quark currents between $\bar K^0$ and
$K^0$ states. Using the vacuum insertion we obtain: $$ < K^0
\mid\bar s\gamma_\alpha(1+\gamma_5)d \bar s
\gamma_\alpha(1+\gamma_5)d \mid \bar K^0 > = $$ $$ = \frac{8}{3}
B_K < K^0 \mid\bar s \gamma_\alpha(1+\gamma_s)d \mid 0
> \times \eqno(8.9) $$ $$ < 0 \mid \bar s
\gamma_\alpha(1+\gamma_5) d\mid \bar K^0
> = -\frac{8}{3} B_K f_K^2 m_K^2 \;\; , $$

where $B_K =1$ if the vacuum insertion saturates this matrix
element.

With the help of eq.(6.5) we obtain: $$m_S - m_L -
\frac{i}{2}(\Gamma_S - \Gamma_L) = 2[Re M_{12} -
\frac{i}{2}\Gamma_{12}] \;\; , \eqno(8.10) $$ where $S$ and $L$
are the abbreviations for $K_S$ and $K_L$, short and long-lived
neutral $K$-mesons respectively. For the difference of masses from
(8.7 - 8.10) we get: $$ m_L -m_S \equiv \Delta m_{LS} =
\frac{G_F^2 B_K f_K^2 m_K}{6\pi^2} \eta_1 m_c^2 V_{cs}^2
V_{cd}^{*^2} \;\; . \eqno(8.11) $$

Constant $f_K$ is known from $K \to l\nu$ decays, $f_K = 160$ MeV.
Gluon dressing of the box diagrams in 4 quark model in the leading
logarithmic (LO) approximation was calculated in \cite{12},
$\eta_1^{LO} = 0.6$. It appears that the subleading logarithms are
numerically very important \cite{13}, $\eta_1^{NLO} = 1.3 \pm
0.2$, the number which we will use in our estimates. We take $B_K
= 1\pm 0.1$ assuming that the vacuum insertion is good
numerically, though the smaller values of $B_K$ can be found in
literature as well (see \cite{14}).

Experimentally the difference of masses is: $$ \Delta m_{LS}^{\rm
exp} = 0.5303(9) \cdot 10^{10} \; {\rm sec}^{-1} \;\; ;
\eqno(8.12) $$

Substituting the numbers in eq. (8.11) we get: $$ \frac{\Delta
m_{LS}^{\rm theor}}{\Delta m_{LS}^{\rm exp}} = 0.6 \pm 0.2 \;\; ,
\eqno(8.13) $$ and we almost get an experimental number from the
short-distance contribution described by the box diagram with
$c$-quarks. As $V_{cs}$ and $V_{cd}$ are already known nothing new
for CKM matrix elements can be extracted from $\Delta m_{LS}$.

Concerning the neutral kaon decays we have: $$\Gamma_S -\Gamma_L =
2\Gamma_{12} \approx \Gamma_S = 1.1 \cdot 10^{10} \; {\rm
sec}^{-1} \;\; (\Delta m_{LS} \approx \Gamma_S/2) \;\; ,
\eqno(8.14) $$ since $ \Gamma_L \ll \Gamma_S$, $\Gamma_L = 2 \cdot
10^7 \; {\rm sec}^{-1}$. $K_L$ is so long-lived because it can
decay only into 3 particles finite states (neglecting CPV) and for
the decays into 3 pions the energy release is small: $$ m_K -
m_{3\pi} = 490 \; {\rm MeV} \;- 3 \cdot 140 \; {\rm MeV} \; = 70
\; {\rm MeV} \;\; . \eqno(8.15) $$

$K_S$ rapidly decays to two pions which have  CP$=+1$.

\section{CPV in $\mbox{\boldmath$K^0 - \bar K^0: K_L \to 2\pi
\; , \;\; \varepsilon_K$}$-hyperbola}

CPV in $K^0 - \bar K^0$ mixing is proportional to the deviation of
$\mid q/p\mid$ from one; so let us calculate this ratio according
to eq. (6.6) taking into account that $\Gamma_{12}$ is real, while
$M_{12}$ is mostly real: $$ \frac{q}{p} = 1-\frac{i Im
M_{12}}{M_{12}-\frac{i}{2}\Gamma_{12}} = 1+ \frac{2i Im
M_{12}}{m_L - m_S +\frac{i}{2}\Gamma_S} \;\; . \eqno(9.1) $$

In this way for the quantity $\tilde\varepsilon$ introduced in eq.
(6.14)
 we obtain: $$ \tilde\varepsilon = -\frac{i Im
M_{12}}{\Delta m_{LS} + \frac{i}{2}\Gamma_S} \;\; . \eqno(9.2) $$

Branching of CP-violating $K_L \to 2\pi$ decay equals: $$ Br (K_L
\to 2\pi^0) + Br (K_L \to \pi^+ \pi^-) = \frac{\Gamma(K_L \to
2\pi)}{\Gamma_{K_L}} = \frac{\Gamma_{K_L \to 2\pi}}{\Gamma_{K_S
\to 2\pi}} \frac{\Gamma(K_S)}{\Gamma(K_L)} = $$ $$ =
\frac{\mid\eta_{00}\mid^2 \Gamma(K_S \to 2\pi^0) +
\mid\eta_{+-}\mid^2 \Gamma(K_S \to \pi^+ \pi^-)}{\Gamma(K_S \to
2\pi^0) +\Gamma(K_S\to \pi^+ \pi^-)}
\frac{\Gamma(K_S)}{\Gamma(K_L)} \approx  $$ $$ \approx
\mid\eta_{00}\mid^2 \frac{\Gamma(K_S)}{\Gamma(K_L)} \approx
\mid\varepsilon\mid^2 \frac{\Gamma(K_S)}{\Gamma(K_L)} \approx
\eqno(9.3) $$ $$\approx \mid\tilde\varepsilon\mid^2 \left |
1-\sqrt 2 \frac{\varepsilon^\prime}{\tilde\varepsilon} \frac{Re
A_0}{Re A_2} \frac{(1+i)}{\sqrt 2}\right |^2 \frac{5.17(4) \cdot
10^{-8}\; {\rm sec}}{0.894(1) \cdot 10^{-10}\; {\rm sec}} \approx
$$ $$ \approx 578(1-0.08) \mid\tilde\varepsilon \mid^2 = 3.02(3)
\cdot 10^{-3} \;\; , $$ where the last number is the sum of $K_L
\to \pi^+ \pi^-$ and $K_L \to \pi^0 \pi^0$ branching ratios and
experimental values of the $K_S$ and $K_L$ widths are used. We
also use the approximate relation $\delta_0 - \delta_2 \approx
45^0$ which follows from the analysis of $\pi - \pi$ scattering.
In eq. (9.3) the small factors of the order of
$\varepsilon^\prime/\tilde\varepsilon$ are neglected while the
enhanced term $\frac{Re A_0}{Re A_2}
\frac{\varepsilon^\prime}{\tilde\varepsilon} = 22.2
\frac{\varepsilon^\prime}{\tilde\varepsilon}$ is taken into
account ($\varepsilon^\prime/\tilde\varepsilon$ is almost real).
This term originates from direct CPV in kaon decays, see the next
section. We used the following approximate formula: $\eta_{00} =
\tilde\varepsilon + i\frac{Im A_0}{Re A_0}$, and estimated the
ratio of amplitudes using eq. (10.10), in which only induced by
QCD penguin term $\sim Im A_0/Re A_0$ is taken into account. The
electroweak penguins partially cancel QCD penguin; omitting them
we underestimate $\mid\tilde\varepsilon\mid$ a bit.

In this way the experimental value of $\mid\tilde\varepsilon\mid$
is determined, and for our theoretical result described by eq.
(9.2) we should have: $$ \mid\tilde\varepsilon\mid = \frac{\mid Im
M_{12}\mid}{\sqrt 2 \Delta m_{LS}} = 2.38(1) \cdot 10^{-3} \;\;
\eqno(9.4) $$

To calculate $Im M_{12}$ one should use eq. (8.8) and the
expression for ${\cal L}_{\Delta S =2}^{eff}$ which follows from
the calculation of the box diagrams. As we already demonstrated
($tt$) box gives the main contribution in $Im M_{12}$. For the
first time it was calculated explicitly not supposing that $m_t
\ll m_W$ in the paper \cite{15a} (a year later the same result was
independently obtained in \cite{15b}): $$ Im M_{12} = -\frac{G_F^2
B_K f_K^2 m_K}{12\pi^2} m_t^2 \eta_2 Im(V_{ts}^2 V_{td}^{*^2})
\times I(\xi) \;\; ,  $$ $$ I(\xi) = \left\{\frac{\xi^2 -11\xi
+4}{4(\xi -1)^2} -\frac{3\xi^2 \ln\xi}{2(1-\xi)^3} \right\} \; ,
\;\; \xi =\left(\frac{m_t}{m_W}\right)^2 \;\; , \eqno(9.5) $$
where factor $\eta_2$ which takes into account the gluon exchanges
in the box diagram with ($tt$) quarks was found in the same paper
\cite{15a} in the leading logarithmic approximation:
$\eta_2^{LO}=0.6$. This factor is not changed substantially by
subleading logs \cite{13}: $\eta_2^{NLO} = 0.57(1)$. Concerning
factor $\eta_3$ which is responsible for gluon dressing of ($ct$)
box and which was as well found in \cite{15a}: $\eta_3^{LO} =
0.4$, the subsequent approximation is \cite{13}: $\eta_3^{NLO} =
0.47(4)$.

Let us present the numerical values for the expression in figure
brackets in eq. (9.5) for several values of the top quark mass
(this factor was obtained in paper \cite{15a} and later in
literature was named Inami-Lim factor, an evident example of the
famous Arnold principle: ``If a notion bears a personal name, then
this name is not the name of the discoverer''): $$ \left\{ \;\;
\right\}
=
\begin{array}{cl} 1 \; , & m_t =0 \; , \;\; \xi =0
\\ 0.55 \; , & \xi =4.7 \; , \;\; {\rm which ~~ corresponds ~~ to}
\;\; m_t = 175 \; {\rm GeV}
\\ 0.25 \; , & m_t = \xi = \infty \end{array} \eqno(9.6) $$

It is clearly seen from (9.5) and (9.6) that the top contribution
to the box diagram is not decoupled (it does not vanish) in the
limit $m_t \to \infty$. Three years earlier the analogous
non-decoupling of $t$-quark contribution through loops to the
quantity $\rho= \frac{\bar g^2/M_Z^2}{g^2/M_W^2}$ was observed in
\cite{16}.

One can easily get where this enhanced at $m_t \to\infty$
behaviour originates from the estimating box diagram in
't~Hooft-Feynman gauge. In the limit $m_t \gg m_W$ the diagram
with two charged higgs exchanges shown in Fig.~6 dominates, since
each vertex of higgs boson emission is proportional to $m_t$.

\begin{figure}[!htb]
\centering \epsfig{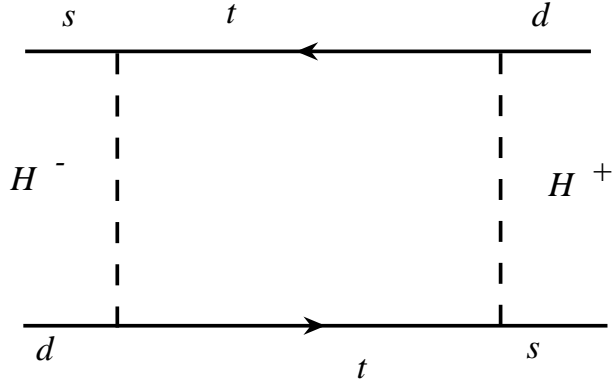} \caption{\em Box
diagram the contribution of which is enhanced as $m^2_t$ in the
limit $m_t \gg m_W$ } \label{WW6Fermi}
\end{figure}

 For the factor which multiplies the four-quark operator from the diagram
shown in Fig.~6 we get: $$ \sim (\frac{m_t}{v})^4 \int\frac{d^4
p}{p^2 -M_W^2)^2} \left[\frac{\hat p}{p^2 -m_t^2}\right]^2 \sim
(\frac{m_t}{v})^4 \frac{1}{m_t^2} = G_F^2 m_t^2 \;\; , \eqno(9.7)
$$ where $v$ is the Higgs boson expectation value.

Substituting eq. (9.5) into (9.4) and substituting the numbers we
obtain: $$ \mid\tilde\varepsilon\mid_{\rm theor} = 0.0075
\bar\eta(1-\bar\rho)(1\pm 0.1) \;\; , \eqno(9.8) $$ $$
\bar\eta(1-\bar\rho) = 0.32(3) \;\; , $$ where 10\% uncertainty in
the value of $B_K = 1 \pm 0.1$ dominates in the error. Taking into
account ($ct$) and ($cc$) boxes we get the following equation:
$$\bar\eta(1.52 -\bar\rho) = 0.32(3) \;\; , \eqno(9.9) $$ which
determines the hyperbola in Fig. 1.

Let us make two comments before finishing this section.
\begin{description}
\item{a)} Why is $\varepsilon_K$  small? From eqs. (9.4), (9.5)
and (8.11) we obtain the following estimate for $\varepsilon_K$:
$$ \varepsilon_K \sim \frac{m_t^2 \lambda^{10} \eta(1-\rho)}{m_c^2
\lambda^2} \;\; . \eqno(9.10) $$

It means that $\varepsilon_K$ is small not because CKM phase is
small, but because $2\times 2$ part of CKM which describes the
mixing of the first two generations is almost unitary and the
third generation almost decouples. We are lucky that the top quark
is so heavy; for $m_t \sim 10$ GeV CPV would not be discovered up
to now (the last statement is true only if the mixing angles of
the first two generations with the third one are not proportional
to $1/\sqrt{m_t}$ -- in the opposite case $\varepsilon_K$ would
not depend on $m_t$ and its smallness would not have any
qualitative explanation).
\item{b)} If the masses of any 2 up (or down) quarks are equal,
then CPV  escapes from CKM matrix. Indeed, for $m_c = m_t$ we
have: $$ Im M_{12} \sim Im [V_{ts} V_{td}^* + V_{cs} V_{cd}^* ]^2
= Im [V_{us} V_{ud}^* ]^2 = 0  \;\; . \eqno(9.11) $$
\end{description}

\section{Direct CPV in $\mbox{\boldmath$K$}$ decays,
$\mbox{\boldmath$\varepsilon^\prime \neq 0 \; (\mid\frac{\bar
A}{A}\mid \neq 1)$}$}

Let us consider the neutral kaon decays into two pions. It is
convenient to deal with the amplitudes of the decays into the
states with definite isospin: $$A(K^0 \to \pi^+ \pi^-) =
\frac{a_2}{\sqrt 3} e^{i\xi_2}e^{i\delta_2} + \frac{a_0}{\sqrt 3}
\sqrt{2} e^{i\xi_0} e^{i\delta_0} \;\; , \eqno(10.1) $$ $$A(\bar
K^0 \to \pi^+ \pi^-) = \frac{a_2}{\sqrt 3}
e^{-i\xi_2}e^{i\delta_2} + \frac{a_0}{\sqrt 3} \sqrt{2}
e^{-i\xi_0} e^{i\delta_0} \;\; , \eqno(10.2) $$ $$A(K^0 \to \pi^0
\pi^0) = \sqrt{\frac{2}{3}} a_2 e^{i\xi_2}e^{i\delta_2} -
\frac{a_0}{\sqrt 3} e^{i\xi_0} e^{i\delta_0} \;\; , \eqno(10.3) $$
$$A(\bar K^0 \to \pi^0 \pi^0) = \sqrt{\frac{2}{3}} a_2
e^{-i\xi_2}e^{i\delta_2} - \frac{a_0}{\sqrt 3} e^{-i\xi_0}
e^{i\delta_0} \;\; , \eqno(10.4) $$ where ``2'' and ``0'' are the
values of ($\pi\pi$) isospin, $\xi_{2,0}$ are the weak phases
which originate from CKM matrix and $\delta_{2,0}$ are the strong
phases of $\pi\pi$-rescattering. If the only quark diagram
responsible for $K\to 2\pi$ decays were charged current tree
diagram which describes $s\to u\bar u d$ transition through
$W$-boson exchange, then the phases would be zero and it would be
no CPV in the decay amplitudes (the so-called direct CPV). All CPV
would originate from $K^0 - \bar K^0$ mixing. Such indirect CPV
was called superweak. However in Standard Model CKM phase
penetrates into the amplitudes of $K\to 2\pi$ decays through the
so-called ``penguin'' diagram shown in Fig. 7, and $\xi_{0,2}$ are
nonzero leading to direct CPV as well.

\begin{figure}[!htb]
\centering \epsfig{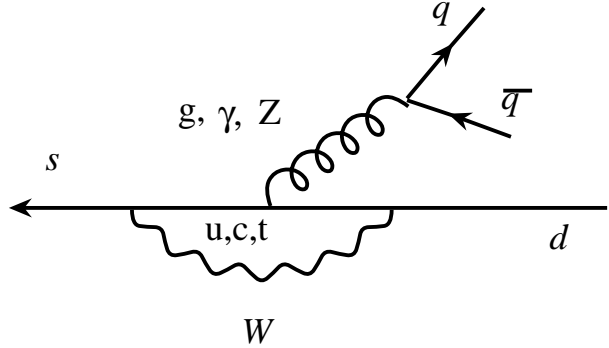} \caption{\em Penguin
diagram responsible for direct CPV in Standard Model }
\label{WW7Fermi}
\end{figure}

From eqs. (10.1) and (10.2) we get: $$\Gamma(K^0 \to \pi^+ \pi^-)
-\Gamma(\bar K^0 \to \pi^+ \pi^-) = -4\frac{\sqrt 2}{3} a_0 a_2
\sin(\xi_2 -\xi_0)\sin(\delta_2 -\delta_0) \;\; ,  $$ so for
direct CPV to occur through the difference of $K^0$ and $\bar K^0$
widths at least two decay amplitudes with different CKM and strong
phases should exist.

In the decays of $K_L$ and $K_S$ mesons violation of CP occurs due
to that in mixing (indirect CPV) and in decay amplitudes of $K^0$
and $\bar K^0$ (direct CPV). The first effect is taken into
account in the expression for $K_L$ and $K_S$ eigenvectors through
$K^0$ and $\bar K^0$. With the help of eq. (6.15) we readily
obtain: $$ K_S = \frac{K^0 +\bar K^0}{\sqrt 2} + \tilde\varepsilon
\frac{K^0 - \bar K^0}{\sqrt 2} \;\; ,  $$ $$ K_L =
\frac{K^0 -\bar K^0}{\sqrt 2} + \tilde\varepsilon \frac{K^0 + \bar
K^0}{\sqrt 2} \;\; , \eqno(10.5) $$ where we neglect $\sim
\varepsilon^2$ terms. For the amplitudes of $K_L$ and $K_S$ decays
into $\pi^+ \pi^-$ with the help of eqs. (10.1), (10.2) we obtain:
$$ A(K_L \to\pi^+ \pi^-) = \frac{1}{\sqrt 2} \left[
\frac{a_2}{\sqrt 3} e^{i\delta_2} 2i\sin\xi_2 + \frac{a_0}{\sqrt
3} \sqrt 2 e^{i\delta_0} 2i\sin\xi_0 \right] + $$ $$ +
\frac{\varepsilon}{\sqrt 2} \left[ \frac{a_2}{\sqrt 3}
e^{i\delta_2} 2\cos\xi_2 + \frac{a_0}{\sqrt 3} \sqrt 2
e^{i\delta_0} 2\cos\xi_0 \right] \;\; , \eqno(10.6) $$ $$ A(K_S
\to\pi^+ \pi^-) = \frac{1}{\sqrt 2} \left[ \frac{a_2}{\sqrt 3}
e^{i\delta_2} 2\cos\xi_2 + \frac{a_0}{\sqrt 3} \sqrt 2
e^{i\delta_0} 2\cos\xi_0 \right] \;\; , \eqno(10.7) $$ where in
the last equation we omit the terms which are proportional to the
product of two small factors, $\varepsilon$ and $\sin\xi_{0,2}$.
For the ratio of these amplitudes we get: $$ \eta_{+-} =
\frac{A(K_L \to\pi^+\pi^-)}{A(K_S \to\pi^+ \pi^-)} =
\tilde\varepsilon + i\frac{\sin\xi_0}{\cos\xi_0} +\frac{i
e^{i(\delta_2 - \delta_0)}}{\sqrt 2} \frac{a_2 \cos\xi_2}{a_0
\cos\xi_0} \left[ \frac{\sin\xi_2}{\cos\xi_2} -
\frac{\sin\xi_0}{\cos\xi_0}\right] \;\; , \eqno(10.8) $$ where we
neglect the terms of the order of $(a_2/a_0)^2 \sin\xi_{0,2}$
because from the $\Delta T = 1/2$ rule in $K$-meson decays it is
known that $a_2/a_0 \approx 1/22$.

The analogous treatment of $K_{L,S} \to \pi^0 \pi^0$ decay
amplitudes leads to: $$ \eta_{00} = \frac{A(K_L
\to\pi^0\pi^0)}{A(K_S \to \pi^0 \pi^0)} = \tilde\varepsilon +
i\frac{\sin\xi_0}{\cos\xi_0} - ie^{i(\delta_2 - \delta_0)} \sqrt 2
\frac{a_2 \cos\xi_2}{a_0 \cos\xi_0}
\left[\frac{\sin\xi_2}{\cos\xi_2} - \frac{\sin\xi_0}{\cos\xi_0}
\right] \;\; . \eqno(10.9) $$

The difference of $\eta_{\pm}$ and $\eta_{00}$ is proportional to
$\varepsilon^\prime$: $$\varepsilon^\prime \equiv \frac{i}{\sqrt
2} e^{i(\delta_2 - \delta_0)} \frac{a_2 \cos\xi_2}{a_0 \cos\xi_0}
\left[\frac{\sin\xi_2}{\cos\xi_2} - \frac{\sin\xi_0}{\cos\xi_0}
\right] = \eqno(10.10) $$ $$ = \frac{i}{\sqrt 2} e^{i(\delta_2 -
\delta_0)} \frac{Re A_2}{Re A_0} \left[ \frac{Im A_2}{Re A_2} -
\frac{Im A_0}{Re A_0} \right] = \frac{i}{\sqrt 2} e^{i(\delta_2
-\delta_0)} \frac{1}{Re A_0} \left[ Im A_2 - \frac{1}{22} Im A_0
\right] \;\; , $$ where $A_{2,0} \equiv e^{i\xi_{2,0}} a_{2,0}$.

Introducing quantity $\varepsilon$ according to the standard
definition: $$\varepsilon = \tilde\varepsilon + i \frac{Im A_0}{Re
A_0} \;\; , $$ we obtain: $$ \eta_{+-} = \varepsilon +
\varepsilon^\prime \; , \;\; \eta_{00} = \varepsilon -
2\varepsilon^\prime \;\; . \eqno(10.11) $$

The double ratio $\eta_{\pm}/\eta_{00}$ was measured in the
experiment and its difference from 1 demonstrates direct CPV in
kaon decays: $$
\left(\frac{\varepsilon^\prime}{\varepsilon}\right)^{\rm exp} =
(1.8 \pm 0.4) 10^{-3} \;\; . \eqno(10.12) $$

The smallness of this ratio is due to (1) the smallness of the
phases produced by the penguin diagrams shown in Fig. 7 and (2)
smallness of the ratio of $a_2/a_0 \approx Re A_2/Re A_0$.

Let us estimate the value of $\varepsilon^\prime$. The penguin
diagram with the gluon exchange generates $K\to 2\pi$ transition
with $\Delta I = 1/2$; those with $\gamma$- and $Z$-exchanges
contribute to $\Delta I = 3/2$ transitions as well. The
contribution of electroweak penguins being smaller by the ratio of
squares of coupling constants is enhanced by the factor $Re A_0/Re
A_2 = 22$, see the last part in eq. (10.10). As a result a partial
compensation of QCD and electroweak penguins occur. In order to
obtain an order of magnitude estimate let us take into account
only QCD penguins. From eq. (10.10) we obtain the following
estimate for the sum of the loops with $t$- and $c$-quarks: $$
\mid\varepsilon^\prime\mid \approx \frac{1}{22\sqrt{2}}
\frac{\sin\xi_0}{\cos\xi_0} =
\frac{1}{22\sqrt{2}}\frac{\alpha_s(m_c)}{12\pi}
\ln(\frac{m_t}{m_c})^2 A^2 \lambda^5  \approx $$ $$\approx 10^{-5}
\frac{\alpha_s(m_c)}{12\pi} \ln (\frac{m_t}{m_c})^2 \;\; .
\eqno(10.13) $$

Taking into account that $\mid\varepsilon\mid \approx 2.4 \cdot
10^{-3}$ we see that the smallness of the ratio of
$\varepsilon^\prime/\varepsilon$ can be readily understood.

In order to make an accurate calculation of
$\varepsilon^\prime/\varepsilon$ one should know the matrix
elements of the quark operators between $K$-meson and two
$\pi$-mesons which with the present knowledge of low-energy QCD is
not possible. That is why a horizontal strip in Fig. 1 which
should correspond to eq. (10.13) has too large width and is not
shown. Nevertheless we discussed direct CPV in this section since
it will be important for $B$-mesons. For more details about
$\varepsilon^\prime/\varepsilon$ estimations see review papers
\cite{17}.

\section{$\mbox{\boldmath$B_d^0 - \bar B_d^0$}$ and $\mbox{\boldmath$B_s^0
- \bar B_s^0$}$ mixing -- two circles}

The $B$-meson semileptonic decays are induced by a semileptonic
$b$-quark decay, $b\to l^- \nu c (l^- \nu u)$. In this way in the
decays of $\bar B^0$ mesons $l^-$ are produced, while in the
decays of $\bar B^0$ mesons $l^+$ are produced. However $B^0$ and
$\bar B^0$ are not the mass eigenstates and being produced at
$t=0$ they start to oscillate according to the following formulas:
$$ B^0(t) = \frac{e^{-i\lambda_+ t} + e^{-i\lambda_- t}}{2} B^0 +
\frac{q}{p} \frac{e^{-i\lambda_+ t} - e^{-i\lambda_- t}}{2} \bar
B^0 \;\; , \eqno(11.1) $$ $$  \bar B^0(t) = \frac{e^{-i\lambda_+
t} + e^{-i\lambda_- t}}{2} \bar B^0 + \frac{p}{q}
\frac{e^{-i\lambda_+ t} - e^{-i\lambda_- t}}{2} B^0 \;\; ; $$ for
derivation use eq. (6.5 - 6.6).

That is why in their semileptonic decays the ``wrong sign
leptons'' are sometimes produced, $l^-$ in the decays of the
particles born  as $B^0$ and $l^+$ in the decays of the particles
born as $\bar B^0$. The amount of these ``wrong sign'' events
depends on the ratio of the oscillation frequency $\Delta m$ and
$B$-meson lifetime $\Gamma$ (unlike the case of $K$-mesons for
$B$-mesons $\Delta\Gamma \ll \Gamma$, see below). For $\Delta
m/\Gamma \gg 1$ a large number of oscillations occurs, and the
number of ``the wrong sign leptons'' equals that of a normal sign.
If $\Delta m \ll \Gamma$, then $B$-mesons decay before they start
to oscillate, and the number of ``the wrong sign events'' is power
suppressed. The pioneering detection of ``the wrong sign events''
by ARGUS collaboration \cite{18} in 1987 demonstrates that $\Delta
m$ is of the order of $\Gamma$, which in the framework of Standard
Model could be understood only if a top quark is unusually heavy,
$m_t \ga 100$ GeV. Fast $B^0 - \bar B^0$ oscillations made
possible the construction of asymmetric $B$-factories where CPV in
$B^0$ decays was observed. (At the end of this introduction let us
mention that UA1 collaboration saw the events which were
interpreted as a possible manifestation of $B_s^0 - \bar B_s^0$
oscillations \cite{19}.)

Integrating the probabilities of $B^0$ decays in $l^+$ and $l^-$
over $t$, we obtain for ``the wrong sign lepton'' probability: $$
W_{B^0 \to \bar B^0} \equiv \frac{N_{B^0 \to l^- x}}{N_{B^0 \to
l^- x} +N_{B^0 \to l^+ x}} = $$ $$ = \frac{\mid \frac{q}{p}\mid^2
(\frac{\Delta m}{\Gamma})^2}{2+(\frac{\Delta m}{\Gamma})^2 +
\mid\frac{q}{p}\mid^2(\frac{\Delta m}{\Gamma})^2} \;\; ,
\eqno(11.2) $$ where we neglect $\Delta\Gamma$, the difference of
$B_+$- and $B_-$-mesons lifetimes. Precisely according to our
discussion for $\Delta m/\Gamma \gg 1$ we have $W = 1/2$, while
for $\Delta m/\Gamma \ll\ 1$ we have $W = 1/2 (\Delta m/\Gamma)^2$
(with high accuracy $\mid p/q\mid =1$, see below).

For $\bar B^0$ decays we get the same formula with the interchange
of $q$ and $p$: $$ W_{\bar B^0 \to B^0} \equiv \frac{N_{\bar B^0
\to l^+ x}}{N_{\bar B^0 \to l^+ x} +N_{\bar B^0 \to l^- x}} = $$
$$ = \frac{\mid \frac{p}{q}\mid^2 (\frac{\Delta
m}{\Gamma})^2}{2+(\frac{\Delta m}{\Gamma})^2 +
\mid\frac{p}{q}\mid^2(\frac{\Delta m}{\Gamma})^2} \;\; .
\eqno(11.3) $$

In ARGUS experiment $B$-mesons were produced in $\Upsilon(4S)$
decays: $\Upsilon(4 S) \to B \bar B$. For $\Upsilon$ resonances
$J^{PC} = 1^{--}$, that is why (pseudo)scalar $B$-mesons are
produced in $P$-wave. It means that $B\bar B$ wave function is
antisymmetric at the interchange  of $B$ and $\bar B$. This fact
forbids the configurations in which due to $B - \bar B$
oscillations both mesons became $B$, or both became $\bar B$.
However after one of the $B$-meson decays the flavor of the
remaining is tagged, and it oscillates according to eqs. (11.1).

If the first decay is semileptonic with $l^+$ emission indicating
that a decaying particle was $B^0$, then the second particle was
initially $\bar B^0$. With the help of eqs. (11.2 - 11.3) and
taking $\mid p/q\mid =1$ we get for the relative number of the
same sign dileptons born in semileptonic decays of $B$-mesons,
produced in $\Upsilon(4S) \to B \bar B$ decays: $$ \frac{N_{l^+
l^+} +N_{l^- l^-}}{N_{l^+ l^-}} =\frac{W}{1-W} = \frac{x^2}{2+
x^2} \; , \;\; x \equiv \frac{\Delta m}{\Gamma} \;\; . \eqno(11.4)
$$

Let us note that if $B^0$ and $\bar B^0$ are produced incoherently
(say, in hadron collisions) a different formula should be used: $$
\frac{N_{l^+ l^+} +N_{l^- l^-}}{N_{l^+ l^-}} =\frac{2W -
2W^2}{1-2W + 2W^2} = \frac{x^2(2+ x^2)}{2+ 2 x^2 + x^4} \;\; .
\eqno(11.5) $$

In the absence of oscillations ($x=0$) both (11.4) and (11.5) are
zero; for high frequency oscillations ($x \gg 1$) both of them are
one.

From the time integrated data of ARGUS and CLEO, the
time-dependent analysis  of $B$-decays at the high energy
colliders (LEP II, Tevatron, SLC) and the time-dependent analysis
at the asymmetric $B$-factories Belle and BaBar the following
result was obtained \cite{6}: $$ x_d = 0.755 \pm 0.015 \;\; .
\eqno(11.6) $$ By using the life time of $B_d$-mesons \cite{6}:
$\Gamma_{B_d} = [1.54(1) \cdot 10^{-12} \; {\rm sec}]^{-1} \equiv
[1.54(1) {\rm ps}]^{-1}$ we get for the mass difference of $B_d$
mesons
 \cite{6}: $$ \Delta m_{B_d} = 0.489(8) {\rm ps}^{-1} \;\; .
\eqno(11.7) $$

In Standard Model $B_d - \bar B_d$ transition occurs through the
box diagram shown in Fig. 8.

\begin{figure}[!htb]
\centering
\epsfig{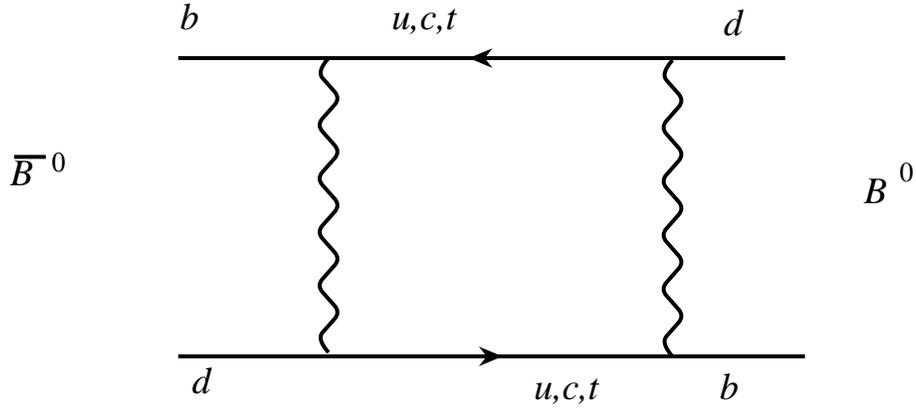}
\caption{\em Diagram responsible for $B_d - \bar B_d$ mixing }
\label{WW8Fermi}
\end{figure}

 Unlike the case of $K^0 - \bar K^0$
transition the power of $\lambda$ is the same for $u$, $c$ and $t$
quarks inside a loop, so the diagram with $t$-quarks dominates.
$M_{12}$ is the same as calculated in \cite{15a, 15b} for the
$K$-meson case.

From eq. (9.5) substituting $B_K f_K^2 V_{ts}^2$ by $B_{B_d} f_B^2
V_{tb}^2$ and removing ``Im'' we obtain: $$ M_{12} = -\frac{G_F^2
B_{B_d} f_B^2}{12\pi^2}m_B m_t^2 \eta_B V_{tb}^2 V_{td}^{*^2}
I(\xi) \;\; , \eqno(11.8) $$ where $I(\xi)$ is the same function
as that for $K$-mesons, eq. (9.5). $\eta_B$ with the account of
NLO corrections is \cite{20}: $$\eta_B^{NLO} = 0.55 \pm 0.01 \;\;
. \eqno(11.9) $$

$\Gamma_{12}$ is determined by the diagram analogous to that shown
in Fig. 4 (where $s$-quark should be substituted by $b$-quark) but
inside the loop $c$-quark can propagate as well (so, 4 diagrams
altogether: $uu$, $uc$, $cu$, $cc$ quarks in the inner lines).
According to \cite{21}: $$ \Gamma_{12} = \frac{G_F^2 B_{B_d} f_B^2
m_B^3}{8\pi} [V_{cb}V_{cd}^*(1+O(\frac{m_c^2}{m_b^2}))
+V_{ub}V_{ud}^*]^2 \;\; , \eqno(11.10) $$ where the term
$O(\frac{m_c^2}{m_b^2})$ accounts for nonzero $c$-quark
mass.\footnote{The dependence of $\Gamma_{12}$ on
$(\frac{m_c^2}{m_b^2})$ is more involved since the contributions
of the diagrams with $uc$ and $cc$ quarks depend differently on
$(m_c/m_b)^2$. However this is not important for what follows.}

Using the unitarity of CKM matrix we get: $$ \Gamma_{12} =
\frac{G_F^2 B_{B_d} f_B^2 m_B^3}{8\pi}[-V_{tb}V_{td}^*
+O(\frac{m_c^2}{m_b^2})V_{cb}V_{cd}^*]^2 \;\; , \eqno(11.11) $$
and the main term in $\Gamma_{12}$ has the same phase as the main
term in $M_{12}$, eq. (11.8). That is why CPV in mixing of
$B$-mesons is suppressed by an extra factor $(m_c/m_b)^2$.
Postponing the discussion of CPV in $B - \bar B$ mixing till the
next section, from eq. (6.5)we obtain: $$ M_+ - M_-
-\frac{i}{2}(\Gamma_+ - \Gamma_-) = 2[\mid M_{12}\mid -
\frac{i}{2}\mid\Gamma_{12}\mid] \;\; , \eqno(11.12) $$ and with
the help of eq. (11.8) for the difference of masses of the two
eigenstates we obtain: $$ \Delta m_{B_d} =-\frac{G_F^2 B_{B_d}
f_B^2}{6\pi^2}m_B m_t^2 \eta_B \mid V_{tb}^2 V_{td}^{*^2}\mid
I(\xi) \;\; , \eqno(11.13) $$ and $\Delta m_{B_d}$ is negative as
well as in the kaon system.

Comparing this theoretical formula with the experimental result
(11.7) we obtain the bound on $\mid V_{td}\mid$, which forms the
circle in Fig.1 with the center at the point $\eta = 0$, $\rho
=1$: $$ \mid V_{td}\mid^2 = A^2 \lambda^6 [(1-\bar\rho)^2 +
\bar\eta^2 ] \;\; . \eqno(11.14) $$

The main uncertainties in this bound on $\bar\eta$, $\bar\rho$
plane are theoretical \cite{6}: $$ B_{B_d} f_{B_d}^2 = (1.3 \pm
0.1) (200 \pm 30 \; {\rm MeV})^2 \;\; , \eqno(11.15) $$ where the
lattice results are used.

Substituting the numbers in (11.13) using (11.7) we get: $$
(1-\bar\rho)^2 + \bar\eta^2 \left|_{_{_{B_d - \bar B_d}}} = (0.85
\pm 0.15)^2 \right. \;\; . \eqno(11.16) $$

If in the diagram shown in Fig. 8 we substitute $s$-quark instead
of $d$-quark we will get $B_s - \bar B_s$ transition. Making
straightforward replacements in eq. (11.13) we obtain: $$ \Delta
m_{B_s} = \Delta m_{B_d} \frac{B_{B_s}f_{B_s}^2}{B_{B_d}f_{B_d}^2}
\mid\frac{V_{ts}^*}{V_{td}^*}\mid^2 \approx  \Delta m_{B_d}(1.2
\pm 0.1)^2 \times $$ $$\times \frac{1}{\lambda^2[(1-\bar\rho)^2
+\bar\eta^2]} = (45 \pm 7) \Delta m_{B_d} = (22 \pm 4) ps^{-1}
\;\; , \eqno(11.17) $$ where for the hadron parameters we use a
lattice result from \cite{6}, the numerical values of $\rho$ and
$\eta$ from the general fit (see Conclusions) and the experimental
result for $\Delta m_{B_d}$. Since the lifetimes of $B_d$~- and
$B_s$~-mesons are almost equal, from (11.6) and (11.17) we get:
$$x_{B_s} = 34 \pm 5 \;\; , \eqno(11.18) $$ which means very fast
oscillations. That is why $W_{B_s}$ equals $1/2$ with very high
accuracy and one cannot extract $x_{B_s}$ from the time integrated
measurements. Performed up to now searches were not sensitive
enough to measure $x_{B_s}$; only the lower bound was obtained
\cite{6}: $$ \Delta m_{B_s} > 13.1 ps^{-1} \;\; {\rm at \; 95\% \;
CL} \;\; . \eqno(11.19) $$

New Tevatron run should be able to find the value of $\Delta
m_{B_s}$ if it is close to Standard Model prediction, eq. (11.17).

However even the lower bound (11.19) appears to be powerful enough
to bound possible values of $\rho$ and $\eta$ quite effectively.
From (11.17) and (11.19) we get: $$ (1-\bar\rho)^2 +\bar\eta^2 <
1.1 \;\; {\rm at \; 95\% \; CL} \;\; , \eqno(11.20) $$ which
produces $\Delta m_{B_s}$ circle in Fig. 1. One can see that this
circle is already inside the outer circle from (11.16)
(recalculated to $2\sigma$ deviation). This progress is possible
because in the ratio $(B_{B_s} f_{B_s}^2)/(B_{B_d} f_{B_d}^2)$
theoretical uncertainty diminishes.

For the difference of the width of $B_{d^+}$ and $B_{d^-}$ we
obtain: $$ \Delta \Gamma_{B_d} = \frac{\mid\Gamma_{12}\mid}{2}
\approx \frac{G_F^2 B_{B_d} f_B^2 m_B^3}{16\pi} \mid V_{td}\mid^2
\;\; , \eqno(11.21) $$ which is very small:
$$\frac{\Delta\Gamma_{B_d}}{\Gamma_{B_d}} \approx 0.3 {\rm \%}
\;\; , \eqno(11.22) $$ as opposite to $K$-meson case, where $K_S$
and $K_L$ lifetimes  differ strongly.

In the $B_s$-meson system a larger time difference is expected;
substituting $V_{ts}$ instead of $V_{td}$ in (11.21) we obtain: $$
\frac{\Delta\Gamma_{B_s}}{\Gamma_{B_s}} \sim 10 {\rm \%} \;\; .
\eqno(11.23) $$

\section{CPV in $\mbox{\boldmath$B^0 - \bar B^0$}$ mixing,
$\mbox{\boldmath$a_{SL}^B$}$ -- too small}

For a long time CPV in $K$-mesons was observed only in $K^0 - \bar
K^0$ mixing. That is why it seems reasonable to start studying CPV
in $B$-mesons from their mixing: $$ \left |\frac{q}{p}\right | =
\left |\sqrt{1+ \frac{i}{2}\left(\frac{\Gamma_{12}}{M_{12}} -
\frac{\Gamma_{12}^*}{M_{12}^*} \right)}\right | = \left | 1+
\frac{i}{4} \left(\frac{\Gamma_{12}}{M_{12}} -
\frac{\Gamma_{12}^*}{M_{12}^*}\right)\right | = $$ $$ =
1-\frac{1}{2} Im \left(\frac{\Gamma_{12}}{M_{12}}\right) \approx 1
- \frac{m_c^2}{m_t^2} Im \frac{V_{cb} V_{cd}^*}{V_{tb} V_{td}^*}
\approx 1- O(10^{-4}) \;\; , \eqno(12.1) $$ where the expressions
for $M_{12}$ (eq. (11.8)) and $\Gamma_{12}$ (eq. (11.11)) were
used. We see that CPV in $B_d - \bar B_d$ mixing is very small
because $t$-quark is very heavy and is even smaller in $B_s - \bar
B_s$ mixing. The experimental observation of $B_d - \bar B_d$
mixing comes from the detection of the same sign leptons produced
in the semileptonic decays of $B_d - \bar B_d$ pair from
$\Upsilon(4S)$ decay. Due to CPV in the mixing the number of $l^-
l^-$ events will differ from that of $l^+ l^+$ and this difference
is proportional to $|\frac{q}{p}| -1  \sim 10^{-4}$. So, one needs
more than $10^8$ semileptonic decays of both $B$ and $\bar B$ to
detect this effect (statistical error $\sim \sqrt N$). Taking into
account that semileptonic branching ratio of $B$ to $e\nu X$ or
$\mu\nu X$ is 20\%, we see that about $10^{10}$ $\Upsilon(4S) \to
B^0 \bar B^0$ decays are needed to observe CPV in mixing according
to the Standard Model. This is completely hopeless now taking into
account that $B$-factories have collected about $10^8$ $B\bar B$
pairs produced in $\Upsilon(4S)$ decays.

Another type of experiment is possible as well. Let us suppose
that in some reactions when $B^+ \bar B^0$ pair is produced $B^+$
is detected, and if $B^- B^0$ is produced $B^-$ is detected. In
this way a flavor state is tagged: it is known what particle
($B^0$ or $\bar B^0$) was produced. If neutral $B$ decays
semileptonically one can look for CPV charge asymmetry: $$
a_{SL}^B = \frac{N_{B^0 \to l^-} -N_{\bar B^0 \to l^+}}{N_{B^0 \to
l^-} + N_{B^0 \to l^+}} = O(10^{-4}) \;\; , \eqno(12.2) $$ while
the experimental result is (in \cite{6} it is denoted by
$a_{CP}$): $$a_{SL}^B = -0.002 \pm 0.009 (\rm stat) \pm 0.008 (\rm
syst) \; \; {\rm -} \eqno(12.3) $$ -- two orders of magnitude
weaker bound on $a_{SL}^B$ than the theoretical prediction.

\section{CPV in interference of mixing and decay
$\mbox{\boldmath$(Im \frac{q \bar A}{pA} \neq 0$}$) }

As soon as it became clear that CPV in $B - \bar B$ mixing is
small theoreticians started to look for another way to find CPV in
$B$ decays. The evident alternative is the direct CPV. It is very
small in $K$-mesons because: a) the third generation almost
decouples in $K$ decays; b) due to $\Delta T = 1/2$ rule. Since in
$B$-meson decays all three quark generations are involved and
there are many different final states one can hope to have large
direct CPV there \cite{22a} - \cite{22c}. An evident drawback of
this strategy: a branching ratio of $B$-meson decays into any
particular exclusive hadronic mode is very small (just because
there are many modes available), so large number of $B$-meson
decays is needed. The specially constructed asymmetric $e^+
e^-$-factories Belle and BaBar collected about thirty million $B
\bar B$ pairs produced in $\Upsilon(4S)$ decays each and
discovered in 2001 CPV in $B(\bar B)$ decays \cite{9'} (at the end
of 2002 statistics is almost 3 times larger in each experiment).

The time evolution of states produced at $t = 0$ as $B^0$ or $\bar
B^0$ is described by eqs. (11.1).

It is convenient to present these formulae in a little bit
different form: $$\mid B^0(t)> = e^{-i\frac{M_+ + M_-}{2}t -
\frac{\Gamma t}{2}}[\cos(\frac{\Delta mt}{2})\mid B^0 >
+i\frac{q}{p}\sin(\frac{\Delta mt}{2})\mid \bar B^0 >] $$
$$\mid\bar B^0(t)> = e^{-i\frac{M_+ + M_-}{2}t - \frac{\Gamma
t}{2}}[+i\frac{p}{q}\sin(\frac{\Delta mt}{2})\mid B^0
> + \cos(\frac{\Delta mt}{2})\mid \bar B^0 >] \; ,  \eqno(13.1)$$
where $\Delta m \equiv M_- - M_+ > 0$, and we put $\Gamma_+ =
\Gamma_- = \Gamma$ neglecting their small difference.

Let us consider  a decay in some final state $f$. Introducing the
decay amplitudes according to the following definitions: $$ A_f =
A(B^0 \to f) \; , \;\; \bar A_f = A(\bar B_0 \to f) \;\; , $$ $$
A_{\bar f} = A(B^0 \to \bar f) \; , \;\; \bar A_{\bar f} = A(\bar
B_0 \to \bar f) \;\; , \eqno(13.2) $$ for the decay probabilities
as functions of time we obtain: $$ P_{B^0 \to f}(t) = e^{-\Gamma
t}\mid A_f\mid^2[\cos^2(\frac{\Delta mt}{2}) +\left |\frac{q\bar
A_f}{pA_f}\right |^2\sin^2(\frac{\Delta mt}{2})
-Im\left(\frac{q\bar A_f}{pA_f}\right)\sin(\Delta mt)] \;\; , $$
$$ P_{\bar B^0 \to \bar f}(t) = e^{-\Gamma t}\mid \bar A_{\bar
f}\mid^2[\cos^2(\frac{\Delta mt}{2}) +\left |\frac{p A_{\bar f}
}{q\bar A_{\bar f}}\right |^2\sin^2(\frac{\Delta mt}{2})
-Im\left(\frac{p A_{\bar f}}{q\bar A_{\bar f}}\right)\sin(\Delta
mt)]\; . \eqno(13.3) $$

The difference of these two probabilities  signal different types
of CPV: the difference in the first two terms in brackets appears
due to direct CPV; that in the last term -- due to CPV in
interference of $B^0 - \bar B^0$ mixing and decays.

Let $f$ be a CP eigenstate: $\bar f = \eta_f f$, where $\eta_f =
+(-)$ for CP even (odd) $f$. (Two examples of such decays: $B^0
\to J/\Psi K_{S(L)}$ and $B^0 \to \pi^+ \pi^-$ are described by
the quark diagrams shown in Fig. 9. The analogous diagrams
describe $\bar B^0$ decays in the same final states.) The
following equalities can be easily obtained:\footnote{The first
one follows from the following identity: $< f-\eta_f \bar f \mid
\Delta V \mid B^0 > =0$, where $\Delta V$ is the interaction
Hamiltonian responsible for the decay; the second one follows from
the analogous identity for $\bar B^0$.} $$ A_{\bar f} = \eta_f A_f
\; , \;\; \bar A_{\bar f} = \eta_f \bar A_f \;\; . \eqno(13.4) $$

\begin{figure}[!htb]
\centering \epsfig{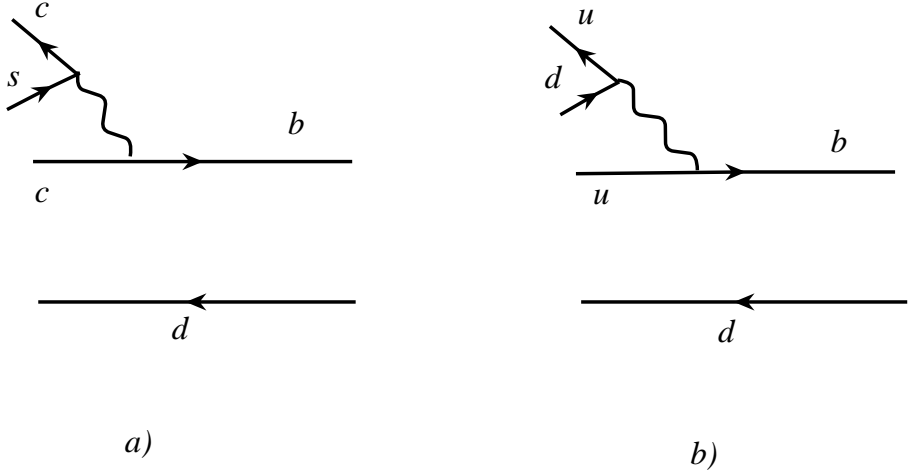} \caption{\em Quark
diagrams describing $B^0 \to J/\Psi K_{S(L)}$ (a)) and $B^0 \to
\pi^+ \pi^-$ (b)) decays } \label{WW9Fermi}
\end{figure}

In the absence of CPV the expressions in brackets in (13.3) are
equal and formulas (13.3) describe the exponential particle decay
without oscillations. Taking CPV into account and neglecting a
small deviation of $\mid p/q\mid$ from one, for CPV asymmetry of
the decays into CP eigenstate we obtain: $$ a_{CP}(t) \equiv
\frac{P_{\bar B^0\to f} - P_{B^0 \to f}}{P_{\bar B^0 \to f} +
P_{B_0 \to f}} = \frac{\mid\lambda\mid^2 -1}{\mid\lambda\mid^2
+1}\cos(\Delta mt) +  \frac{2 Im\lambda}{\mid\lambda\mid^2 +1}
\sin(\Delta mt) \equiv $$ $$ \equiv -C_f \cos(\Delta mt) + S_f
\sin(\Delta mt) \;\; , \eqno(13.5) $$ where $$\lambda \equiv
\frac{q \bar A_f}{p A_f} \;\; . \eqno(13.6) $$ (do not confuse
with the parameter of CKM matrix), $C_f \equiv
\frac{1-\mid\lambda\mid^2}{1+ \mid\lambda\mid^2}$, $S_f \equiv
\frac{2 Im \lambda}{\mid\lambda \mid^2 +1}$.\footnote{Belle
collaboration uses $S_f$ and $A_f \equiv -C_f$.} The nonzero value
of $C_f$ corresponds to direct CPV; it occurs when more than one
amplitude contribute to the decay. For extraction of CPV
parameters (angles of unitarity triangle) in this case the
knowledge of strong rescattering phases is necessary. The
nonvanishing $S_f$ describes CPV in interference of mixing and
decay. It is nonzero even when there is only one decay amplitude,
and $|\lambda | =1$. Such decays are of special interest since the
extraction of CPV parameters becomes independent on poorly known
strong phases of the final particles scattering.

The decays of $\Upsilon(4S)$ resonance produced in $e^+ e^-$
annihilation  are a powerful source of $B^0 \bar B^0$ pairs. A
semileptonic decay of one of the $B$'s tags ``beauty'' of the
partner at the moment of decay thus making it possible to study
CPV. However the time-integrated asymmetry is zero for decays were
$C_f$ is zero. This happens since we do not know which of the two
$B$-mesons decays earlier, and asymmetry is proportional to: $$ I
= \int\limits^\infty _{-\infty} e^{-\Gamma |t|} \sin (\Delta mt)
dt = 0 \;\; . $$

The asymmetric $B$-factories provide possibility to measure the
time-dependence: $\Upsilon(4S)$ moves in a laboratory system, and
since energy release in $\Upsilon(4S) \to B\bar B$ decay is very
small both $B$ and $\bar B$ move with the same velocity as the
original $\Upsilon(4S)$. This makes the resolution of $B$ decay
vertices possible unlike the case of $\Upsilon(4S)$ decay at rest,
when $B$ and $\bar B$ being non-relativistic decays at almost the
same point (for the detailed description of the experiments see
the lecture of A.~Bondar in these proceedings) \cite{23}. The
implementation of the time-dependent analysis for the search of
CPV in $B$-mesons was suggested in papers \cite{24a, 24b}.

\section{$\mbox{\boldmath$B_d^0(\bar B_d^0) \to J/\Psi K_{S(L)}$}$,
$\mbox{\boldmath$\sin 2\beta$}$ -- straight lines}

Currently this is the only decay channel where both $B$-factories
observe statistically significant CPV.

The tree diagram contributing to this decay is shown in Fig.~9a).
The product of the corresponding CKM matrix elements is: $V_{cb}^*
V_{cs} \simeq A\lambda^2$. Also the penguin diagram $b\to sg$ with
the subsequent $g \to c \bar c$ decay contributes to the decay
amplitude. Its contribution is proportional to: $$ P \sim V_{us}
V_{ub}^* f(m_u) + V_{cs} V_{cb}^* f(m_c) + V_{ts} V_{tb}^* f(m_t)
= $$ $$ = V_{us} V_{ub}^* (f(m_u) - f(m_t)) + V_{cs} V_{cb}^*
(f(m_c) - f(m_t)) \;\; , \eqno(14.1) $$ where function $f$
describes the contribution of quark loop and we subtracted zero
from the expression on the first line just as we did when
calculating the box diagram in Section 8. The last term on the
second line has the same weak phase as the tree amplitude, while
the first term has a CKM factor $V_{us} V_{ub}^* \sim \lambda^4
(\rho - i\eta)A$. Since (one-loop) penguin amplitude should be in
any case smaller than the tree one we get that with 1\% accuracy
there is only one weak amplitude governing $B_d^0 (\bar B_d^0) \to
J/\Psi K_{S(L)}$ decays. This is the reason why this mode is
called a ``gold-plated mode'' -- the accuracy of the theoretical
prediction of the CP-asymmetry is very high, and $Br (B_d \to
J/\Psi K^0) \approx 10^{-3}$ is large enough to detect CPV.
Substituting in Eq.(13.5) $|\lambda | =1$ we obtain: $$a_{CP}(t) =
Im \lambda \sin(\Delta m \Delta t) \;\; , \eqno(14.2) $$ where
$\Delta t$ is the time difference between the semileptonic decay
of one of $B$-mesons produced in $\Upsilon(4S)$ decay and that of
the second one to $J/\Psi K_{S(L)}$. Using the following equation:
$$\bar A_f = \eta_f \bar A_{\bar f} \;\; , \eqno(14.3) $$ where
$\eta_f$ is CP parity of the final state, we obtain: $$\lambda =
\left(\frac{q}{p}\right)_{B_d} \frac{A_{\bar B^0 \to J/\Psi
K_{S(L)}}}{A_{B^0 \to J/\Psi K_{S(L)}}} =
\left(\frac{q}{p}\right)_{B_d} \eta_f \frac{A_{\bar B^0 \to
\overline{J/\Psi K_{S(L)}}}}{A_{B^0 \to J/\Psi K_{S(L)}}} \;\; .
\eqno(14.4) $$

The amplitude in the nominator contains $\bar K^0$ production. To
project it on $\bar K_{S(L)}$ we should use: $$ \overline{K^0} =
\frac{K_S - K_L}{(q)_K} = \frac{\bar K_S + \bar K_L}{(q)_K} \;\; ,
\eqno(14.5) $$ getting $(q)_K$ in denominator. The amplitude in
the denominator contains $K^0$ production, and using: $$ K^0 =
\frac{K_S + K_L}{(p)_K} \eqno(14.6) $$ we obtain the factor
$(p)_K$ in the nominator. Collecting all the factors together and
substituting CKM matrix elements for $\bar A_{\bar f}/A_f$ ratio
we get: $$\lambda = \eta_{S(L)} \left(\frac{q}{p}\right)_{B_d}
\frac{V_{cb} V_{cs}^*}{V_{cb}^* V_{cs}} \left(\frac{p}{q}\right)_K
\;\; . \eqno(14.7) $$

Since in $B$ decays $J/\Psi$ and $K_{S(L)}$ are produced in
$P$-wave, $\eta_{S(L)} = -(+)$ (CP of $J/\Psi$ is $+$, that of
$K_S$ is $+$ as well, and $(-1)^l = -1$ comes from $P$-wave; CP of
$K_L$ is $-$).

Substituting the expressions for $(q/p)_{B_d}$ and $(p/q)_K$ in
(14.4) and taking into account $\eta_{S(L)}$ we obtain: $$
\lambda(J/\Psi K_{S(L)}) = -(+) \frac{V_{td} V_{tb}^*}{V_{td}^*
V_{tb}} \frac{V_{cb} V_{cs}^*}{V_{cb}^* V_{cs}} \frac{V_{cd}^*
V_{cs}}{V_{cd} V_{cs}^*} \;\; , \eqno(14.8) $$ which is invariant
under the phase rotation of any quark field. From eq. (3.20) and
Fig.~2 we have: $$ \arg (V_{tb}^* V_{td}) = 2\pi - \beta \;\; ,
\eqno(14.9) $$ and we finally obtain: $$ a_{CP}(t)\left
|_{_{_{_{\large J/\Psi K_{S(L)}}}}}  = (-) \sin(2\beta)
\sin(\Delta m \Delta t) \right. \;\; , \eqno(14.10) $$ which is a
simple prediction of Standard Model. In this way the measurement
of this asymmetry at $B$-factories provides the value of the angle
$\beta$ of the unitarity triangle. The results of Belle and BaBar
are consistent; their average is: $$ \sin 2\beta = 0.73 \pm 0.05
({\rm stat}) \pm 0.035 ({\rm syst}) \;\; . \eqno(14.11) $$

This result is based on the analysis of approximately $80 \cdot
10^6$ pairs of $B \bar B$ produced in $\Upsilon(4S)$ decays per
collaboration \cite{25}. As a final state not only $J/\Psi
K_{S(L)}$ were selected, but also the other states with hidden
charm
\\($\Psi^\prime K_S, \eta_c K_S, \chi_c K_S$). The value of
$|\lambda |$ was also determined from the absence of $\cos(\Delta
m \Delta t)$ term in asymmetry: $$ |\lambda | = 0.95 \pm 0.035
({\rm stat}) \pm 0.025 ({\rm syst}) \eqno(14.12) $$ in accordance
with Standard Model prediction. From eq. (14.11) we obtain 2
possible solutions with angle $2\beta$ in the first or second
quadrant. Both are shown in Fig.~1 by straight lines and the first
one coincides nicely with Standard Model expectations.

Returning to formula (14.8) let us note that the decay amplitudes
and $K^0 - \bar K^0$ mixing do not contain complex phases, that is
why the only source of it in CP-asymmetry in $B^0 \to J/\Psi K$
decays is $B^0 - \bar B^0$ mixing:

$$ \left(\frac{q}{p}\right)_{B_d} = \sqrt\frac{M_{12}^*}{M_{12}} =
\frac{V_{tb}^* V_{td}}{V_{tb} V_{td}^*} \;\; , \eqno(14.13) $$
thus the phase comes from $V_{td}$, that is why the final
expression (14.10) contains the angle $2\beta$ -- the phase of
$V_{td}/V_{td}^*$.

\section{$\mbox{\boldmath$B_d \to \pi^+ \pi^-, \sin 2\alpha$}$,
Penguin versus tree, $\mbox{\boldmath$\mid\bar A/A\mid \neq 1$}$}

The pair $\pi^+ \pi^-$ produced in $B_d$ decay has positive CP:
$CP(\pi\pi)_{l=0} = +1$. The tree level diagram contributing to
this decay is shown in Fig.~9b. Let us suppose for a moment that
it dominates in the decay probability analogously to $B\to J/\Psi
K$ case. Since CKM matrix element $V_{ub}$ has a nonzero phase the
CP-asymmetry should be different from that in $J/\Psi K$ decays.
Let us calculate it: $$ \lambda = \left(\frac{q}{p}\right)_B
\frac{\bar A}{A} = \frac{V_{td} V_{tb}^*}{V_{td}^* V_{tb}} \cdot
\frac{V_{ub} V_{ud}^*}{V_{ub}^* V_{ud}} = e^{-i(2\beta +2\gamma)}
\;\; , \eqno(15.1) $$ $$ Im \lambda = \sin 2\alpha \;\; , $$ $$
a_{CP}(t)\left|_{_{_{\pi^+ \pi^-}}} = \sin(2\alpha)\sin(\Delta mt)
\right. \;\; $$ where the triangle relation $\alpha + \beta
+\gamma = \pi$ is used. So, the study of $t$-dependent CP
asymmetry in $\Upsilon(4S) \to B \bar B \to l^{\pm} X$ $\pi^+
\pi^-$ decay would measure angle $\alpha$ {\it if} tree diagram
dominates in $B \to \pi\pi$ decay. The penguin diagrams producing
the transition $b \to d g \to d \bar u u $ also contribute to the
$\pi^+ \pi^-$ decay mode: $$ P \sim V_{ud} V_{ub}^* f(m_u) +
V_{cd} V_{cb}^* f(m_c) + V_{td} V_{tb}^* f(m_t) = $$ $$ = V_{ud}
V_{ub}^*[f(m_u) -f(m_t)] + V_{cd} V_{cb}^*[f(m_c) -f(m_t)] \;\; ,
\eqno(15.2) $$ and while the first term should be added to the
tree diagram (CKM phase is the same), the second one has a
different phase and is of the order of $\lambda^3$, just like the
tree diagram. Naively one would expect that the penguin
contribution should be nevertheless much suppressed. Feynman
diagram calculation leads to the following damping factor: $$ P/T
\sim \frac{\alpha_S(m_b)}{12\pi} \ln
\left(\frac{m_t}{m_b}\right)^2 \approx 0.04 \;\; . \eqno(15.3) $$

However  the comparison of $B \to \pi\pi$ and $B \to \pi K$
branching ratios demonstrates that the naive estimate is not valid
and the penguin contributions are very much enhanced. The
amplitudes of $B\to \pi\pi$ decays are of the order of: $$ A_{B\to
\pi\pi} \sim \lambda^3 T + \lambda^3 P \;\; , \eqno(15.4) $$ (here
$\lambda$ is the CKM parameter, $\lambda \approx 0.22$), while
that for $B \to \pi K$ decays are: $$ A_{B\to \pi K} \sim
\lambda^4 T + \lambda^2 P \;\; , \eqno(15.5) $$ and if estimate
(15.3) were correct, the decay probabilities with $K$-meson
production should be suppressed as $\lambda^2 \sim 5 \cdot
10^{-2}$. Experimentally a number of branchings of these two types
of decays were measured, or upper bounds were established
(different decay modes have different charges of $B$, $K$ and
$\pi$). It appears that the decays to $\pi K$ are even more
probable than to $\pi\pi$ \cite{6}. This can happen only for $P/T
> \lambda$; it follows from the experimental data that: $$ (P/T)_{\rm
exp} \sim \sqrt{\lambda} \;\; . \eqno(15.6) $$

Taking into account the penguin contribution instead of eq.~(15.1)
we get \cite{26}: $$ \lambda = \left[e^{2i\alpha}
\frac{1+|\frac{P}{T}| e^{i(\delta +\gamma)}}{1+|\frac{P}{T}|
e^{i(\delta -\gamma)}}\right] \;\; , \eqno(15.7) $$ where the part
of the penguin contribution proportional to $V_{ud} V_{ub}^*$ is
included into the tree amplitude and $\delta \equiv \delta_P -
\delta _T$ is the difference of strong phases of the penguin and
tree amplitudes. $\alpha$ and $\gamma$ are the angles of unitarity
triangle.

The experimental data on CP asymmetries in $B\to \pi^+ \pi^-$
decay are currently controversial. According to Belle \cite{27}:
$$S_{\pi\pi} = -1.23 \pm 0.41 \pm 0.08$$ $$C_{\pi\pi} = -0.77 \pm
0.27 \pm 0.08 \;\; , \eqno(15.8) $$ where the first errors are
statistical, the second -- systematical. Both numbers are
different from zero at the level of 3 sigma. Since $C_{\pi\pi}$ is
different from zero the penguin contributions are not negligible.
Extracting the penguin amplitude from the $B\to \pi K$ decay
branching ratios with the help of flavor SU(3) and the tree
amplitude from the factorization hypothesis applied to $B\to \pi e
\nu$ decay the authors of \cite{28} conclude that $|P/T| \sim
0.3$. Using this result and the triangular relation $\gamma = \pi
-\beta -\alpha$ one can extract the values of $\alpha$ and
$\delta$ from (15.8). The result is \cite{27}: $$ 78^0 < \alpha <
152^0 \eqno(15.9) $$ for $\beta = 23^0$ and $0.15 < |P/T| < 0.45$.

BaBar Collaboration did not observe CP violation in $B_d \to \pi^+
\pi^-$ decay \cite{29}: $$S_{\pi\pi} = 0.02 \pm 0.34 \pm 0.05$$
$$C_{\pi\pi} = - 0.30 \pm 0.25 \pm 0.04 \;\; . \eqno(15.10) $$

We can hope that with more data available and corresponding
diminishing of statistical error the Belle-BaBar controversy on
CPV in $B_d \to \pi^+ \pi^-$ decay will be resolved.

\section{Angle $\mbox{\boldmath$\gamma$}$}

Angle $\beta$ of the unitarity triangle is already measured with
good accuracy; angle $\alpha$ will be determined from CP asymmetry
in $B_d \to \pi^+ \pi^-$ decays with better accuracy when more
statistics will become available. Evidently the next task is to
measure angle $\gamma$, or the phase of $V_{ub}$. In $B_d$ decays
angle $\beta$ enters the game through $B_d - \bar B_d$ mixing. To
avoid it in order to single out the angle $\gamma$ let us turn to
$B_s$ decays: CKM matrix elements $V_{ts}$ and $V_{tb}$ which
participate in $B_s - \bar B_s$ mixing are real.\footnote{Another
way to avoid $B_d - \bar B_d$ mixing is to look for direct CPV in
$B^\pm$ decays.} Selecting the decays where $b \to u$ transition
dominates and looking for CP asymmetry we can measure angle
$\gamma$. $B_s$ decays to CP eigenstates which occur through $b
\to u$ transition would provide the necessary information: $B_s
\to K_S \rho, K_S \pi^0$. However the penguin transition $b \to d
g \to d \bar u u$ is proportional to $\lambda^3$ as well as the
tree decay $b \to u \bar u d$. That is why the extraction of
$\gamma$ from these decays will be as involved as that of $\alpha$
from $B_d \to \pi^+ \pi^-$ decays.

The analogs of ``golden'' $B_d \to J/\Psi K$ mode in the case of
$B_s$ are $B_s \to J/\Psi \phi, J/\Psi \eta^\prime$ decays. In the
framework of Standard Model the CP asymmetries in these decays
should be zero:  CKM phase is absent in $B_s - \bar B_s$ mixing
and in $b \to c \bar c s$ decay amplitude (CP $(J/\Psi
\eta^\prime) = +$, $J/\Psi \phi$ is a mixture of CP odd and even
states, but this fact only dilutes asymmetry). That is why the
search of this asymmetry is very interesting: a nonzero result
discovers New Physics (most probably in $B_s - \bar B_s$ mixing).

An interesting strategy for the measurement of angle $\gamma$ was
suggested in paper \cite{29'}: the study of the time-dependent
decay asymmetries of $\bar B_s \to D_s^+ K^-$, $B_s \to D_s^+ K^-$
as well as $\bar B_s \to D_s^- K^+$, $B_s \to D_s^- K^+$ decays.
The diagrams contributing to the first pair of decays are shown in
Fig.~10. The final states are not CP eigenstates (unlike the
previously discussed $B_d(\bar B_d)$ decays).

\begin{figure}[!htb]
\centering
\epsfig{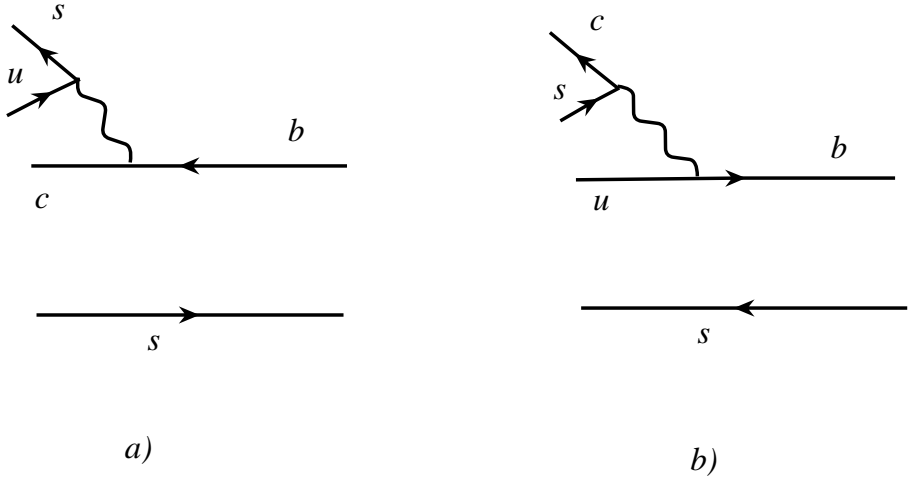}
\caption{\em  Dominant diagrams contributing to  $\bar B_s \to D_s^+ K^-$ (a) and $B_s \to D_s^+ K^-$ (b) decays }
\label{WW10Fermi}
\end{figure}

Tagging a parent meson we can study the time-dependent asymmetry
of $\bar B_s \to D_s^+ K^-$ and $B_s \to D_s^+ K^-$ decays and
extract from it the following quantity: $$ \lambda_{D_s^+ K^-} =
\left(\frac{q}{p}\right)_{B_s} \frac{A_1 V_{us}^* V_{cb}}{A_2
V_{ub}^* V_{cs}} \;\; , \eqno(16.1) $$ see eqs. (13.6), (14.4),
where amplitude $A_1$ corresponds to Fig.~10a, and amplitude $A_2$
corresponds to Fig.~10b (CKM matrix elements are written
separately). Analogously from the study of $\bar B_s \to D_s^-
K^+$, $B_s \to D_s^- K^+$ decays we extract: $$\lambda_{D_s^- K^+}
= \left(\frac{q}{p}\right)_{B_s} \frac{A_2 V_{ub} V_{cs}^*}{A_1
V_{us} V_{cb}^*} \;\; . \eqno(16.2) $$

Multiplying these two lambdas we get: $$\lambda_{D_s^+ K^-} \times
\lambda_{D_s^- K^+} = \left(\frac{q}{p}\right)^2_{B_s}
\frac{V_{ub} V_{cs}^*}{V_{ub}^* V_{cs}} \frac{V_{us}^*
V_{cb}}{V_{us}V_{cb}^*} = e^{-2i\gamma} \;\; , \eqno(16.3) $$ the
unknown hadronic amplitudes $A_i$ cancel, $(q/p)_{B_s}$ is real
and phase $\gamma$ will be measured -- the ``only'' problem is to
collect enough statistics of tagged $B_s(\bar B_s) \to D_s^+ K^-
(D_s^- K^+)$ decays and to measure its time-dependent asymmetries.

Another way to determine $\gamma$ is through $B_d$ decays into $D
K$ final states \cite{29''''}, see also \cite{29'''''}.

Study of $B^\pm$ and $B_d$ decays into $\pi K$ final states also
allows to determine angle $\gamma$ \cite{29''}; the present
experimental uncertainties do not allow to draw definite
conclusions.

Other strategies to measure (or constrain) angle $\gamma$ can be
found in literature; for the list of references look at a recent
review \cite{29'''}.

\section{CPV in $\mbox{\boldmath$B\to \phi K_S, K^+ K^- K_S, \eta^\prime
K_S$}$: penguin domination}

These decays are dominated by penguin diagrams $b \to s g \to s s
\bar{s}$.
 The diagram with an intermediate $u$-quark is
proportional to $\lambda^4$, while those with intermediate $c$-
and $t$-quarks are proportional to $\lambda^2$. In this way the
main part of the decay amplitude is free of CKM phase, just like
in case of $B_d \to J/\Psi K$ decays. A nonzero phase which leads
to time-dependent CP asymmetry comes from $B_d - \bar B_d$
transition: $$ a_{CP}(t) = -\eta_f \sin(2\beta) \sin(\Delta m
\Delta t) \;\; , \eqno(17.1) $$ analogously to $B_d \to J/\Psi K$
decays, eq. (14.10) (\cite{30'}, see also \cite{30''}). $\phi K_S$
and $\eta^\prime K_S$ final states are CP-odd, while in case of
the decay into three kaons final state is a mixture of CP-even and
odd states. According to \cite{30} CP-even final states dominate.
Here are Belle results \cite{30}: $$
\begin{array}{ccc} {\rm Mode} & & \sin 2\beta \\ \phi K_S & & -0.73
\pm 0.64 \pm 0.22 \\ K^+ K^- K_S & & 0.49 \pm 0.43 \pm
0.11^{+0.33}_{-0.00} \\ \eta^\prime K_S & & +0.71 \pm 0.37 \pm
0.05 \;\; ,
\end{array}
\eqno(17.2) $$ where the third error for the $K^+ K^- K_S$ mode
arises from uncertainty in the fraction of the CP-odd component.
The value of $\sin 2\beta$ obtained from CP asymmetry in $B_d \to
\phi K_S$ decay deviates from the Standard Model expectation (eq.
(14.1)) by more than $2\sigma$. According to BaBar \cite{31}: $$
\sin 2\beta(\phi K_S) = -0.19 \pm 0.52 \pm 0.09 \;\; , \eqno(17.3)
$$ and the deviation is more moderate. A number of New Physics
contributions to the penguin diagram were immediately suggested
which explain the deviation from Standard Model in $\phi K_S$ mode
observed by Belle Collaboration. More statistics is needed to
clarify the present situation: perfect SM description of CP
asymmetry in $B_d \to J/\Psi K$ decays which occurs at the tree
level and possible New Physics contribution to loop-induced $B_d
\to \phi K_S$ decay. Belle and BaBar differ not only in $S_{\phi
K_S}$, but in $C_{\phi K_S}$ as well: $$ C_{\phi K_S} ({\rm
Belle}) = 0.56 \pm 0.41 \pm 0.16 $$ $$ C_{\phi K_S} ({\rm BaBar})
= -0.80 \pm 0.38 \pm 0.12 \eqno(17.4) $$

\section{Conclusions: CKM fit and future prospects}

Four parameters of CKM matrix are fitted from the following
experimental data: $|V_{ud}| = 0.9734 \pm 0.0008$ \cite{6},
$|V_{us}| = 0.2196 \pm 0.00261$ \cite{6}, $|V_{cd}| = 0.224 \pm
0.016$ \cite{6}, $|V_{cs}| = 0.996 \pm 0.013$ \cite{6}, $|V_{cb}|
= 0.041 \pm 0.002$ \cite{6}, $|V_{ub}| = 0.0036 \pm 0.0007$
\cite{6}, $\sin 2\beta = 0.73 \pm 0.06$, $|\varepsilon_K| = (2.282
\pm 0.017) \cdot 10^{-3}$, $\Delta M_{B_d} = 0.489 \pm 0.008$
ps$^{-1}$. Here are the results of the fit: $$ \lambda = 0.223 \pm
0.002$$ $$ A = 0.82 \pm 0.04$$ $$ \bar\eta = 0.32 \pm 0.04
\eqno(18.1) $$ $$\bar\rho = 0.24 \pm 0.08$$ $$ \chi^2/n.d.o.f. =
7.8/5 \;\; .$$

The good quality of the fit is caused partly by avoiding
controversial data, like eq. (17.2).

For the angles of a unitarity triangle we obtain: $$ \beta = 23^0
\pm 3^0 $$ $$ \alpha = 103^0 \pm 9^0 \eqno(18.2) $$ $$ \gamma =
54^0 \pm 9^0 $$

Future prospects are:
\begin{enumerate}
\item Diminishing the experimental errors in $\beta$ from $B
\to J/\Psi K$ and in $\alpha$ from $B \to \pi^+ \pi^-$ CPV
asymmetries -- the better accuracy in CKM parameters;
\item Measurement of asymmetries in $B \to \phi K_S$ decay with
better accuracy -- check of SM;
\item Measurement of $\Delta M_{B_s}$ through $B_s - \bar B_s$
oscillation -- check of SM;
\item Measurement of the angle $\gamma$ -- check of Standard Model.
\end{enumerate}

The rare $K$-meson decay $K^+ \to \pi^+ \nu\nu$ and $K^0 \to \pi^0
\nu\nu$ are also sensitive to the values of CKM parameters $\rho$
and $\eta$ which make the measurements of the corresponding decay
probabilities interesting: the deviations from SM predictions
would signal New Physics.

In this way the study of flavor physics in pre-LHC era could put
an end to thirty years success of Standard Model.

I am grateful to L.B.~Okun for fruitful discussions, to
E.A.~Andriash and G.G.~Ovanesyan for performing CKM fit and
drawing Fig.~1, to Ya.I.~Azimov, M.~Gronau, T.~Hambye, A.~Soni and
M.B.~Voloshin for useful remarks. I am grateful to E.A.~Ilyina for
her help in preparing the manuscript.

\end{document}